\documentclass[aps,prd,twocolumn,nofootinbib,eqsecnum,superscriptaddress,longbibliography]{revtex4-2}

\usepackage[centertags]{amsmath}
\usepackage{amssymb}
\usepackage{latexsym}
\usepackage{enumerate}
\usepackage{graphicx}
\usepackage{mathrsfs}
\usepackage[hidelinks]{hyperref}
\usepackage{stmaryrd}
\usepackage{import}
\usepackage{tensor}
\usepackage{xcolor}
\usepackage{bm}
\usepackage{multirow}
\usepackage{breakurl}
\usepackage{float}
\usepackage{lipsum}
\usepackage{siunitx}
\usepackage{comment}
\usepackage{mathtools}
\usepackage{autobreak}
\usepackage{subfig}
\usepackage{orcidlink}
\mathtoolsset{centercolon}
\usepackage{comment}

\newcommand{\dU}{\mathrm{U}}
\newcommand{\dV}{\mathrm{V}}

\newcommand{\dM}{\mathrm{M}}

\newcommand{\dd}{\mathrm{d}}
\newcommand{\de}{\mathrm{e}}
\newcommand{\di}{\mathrm{i}}

\include{preamble}

\allowdisplaybreaks

\begin{document}

\title{Schott term in the binding energy for compact binaries\\  on circular orbits at fourth post-Newtonian order}

\begin{abstract}
    The phasing for compact binary systems on circular orbits was obtained in \mbox{\href{https://doi.org/10.1103/PhysRevLett.131.121402}{[Phys. Rev. Lett. \textbf{131}, 121402 (2023)]}} at fourth-and-a-half post-Newtonian (4.5PN) order thanks to two main ingredients: the 4PN conservative energy (associated to a nonradiative spacetime) in terms of the orbital frequency and the 4.5PN flux in terms of the waveform frequency (i.e., the half-frequency of the $(\ell,m)=(2,2)$ mode). When obtaining the phasing, a key physical postulate was made: the expression of the \textit{binding} energy in terms of the \textit{waveform} frequency was assumed to be identical to the expression of the \textit{conservative} energy in terms of the \textit{orbital} frequency. This postulate was necessary to  ensure that the frequency evolution obtained through the flux-balance law (which involves the binding energy) was independent of the choice of spacetime foliation. In this work, I show that the binding energy entering the flux-balance law differs from the 4PN conservative energy by a 4PN pseudo-Schott term, associated with radiation-reaction effects due to gravitational tails. Unlike the usual Schott terms (at 2.5PN, 3.5PN, and 4.5PN), the pseudo-Schott term is not a total derivative and is in fact hereditary, so it does not vanish for circular orbits. Remarkably, the binding energy thus obtained is in perfect agreement with the one obtained using the aforementioned physical postulate, which confirms that the 4.5PN phasing associated to the waveform frequency computed in \mbox{\href{https://doi.org/10.1103/PhysRevLett.131.121402}{[Phys. Rev. Lett. \textbf{131}, 121402 (2023)]}}  is indeed correct. This result is extended to the other Poincaré invariants, and ``thermodynamic'' relations between the binding energy and angular momentum are established.  Finally, the chirp and phasing associated to the \textit{orbital} frequency are presented at 4.5PN, including horizon-absorption effects.
\end{abstract}

\author{David Trestini\,\orcidlink{0000-0002-4140-0591}}
\affiliation{School of Mathematical Sciences and STAG Research Centre, University of Southampton, Southampton, United Kingdom, SO17 1BJ}
\email{david.trestini@soton.ac.uk}
\date{\today}
\maketitle

\section{Introduction}
\label{sec:intro}
Given the sensitivities of current and future gravitational detectors, it is important to develop extremely precise waveforms for compact binary systems. This effort has been pushed in particular using the post-Newtonian approximation, which is an expansion in the small orbital velocity $v$ of the binary. This expansion describes the entire inspiral phase of the binary, and yields entirely analytical results, where corrections of order $(v/c)^{2n}$ are said to be of $n$-th post-Newtonian ($n$PN) order.

In order to provide gravitational waveform templates, first it is  key to understand the motion of the compact binary: this is described by the equations of motion, which are given as the usual Newtonian acceleration plus post-Newtonian corrections, which depend only on the masses, positions and velocities of the two particles. These can be decomposed into: (i) a conservative, time-symmetric piece; and (ii) a dissipative, time-asymmetric, radiation-reaction piece. The conservative acceleration is currently known at 4PN (the 4.5PN contribution is vanishing) thanks to several groups using various methods: the Arnowitt–Deser–Misner (ADM)  Hamiltonian formalism  \cite{BiniD13, DJS14,Damour:2014jta,JaraS15,DJS15eob, DJS16}, the Fokker action \cite{BBBFMa,BBBFMb,BBBFMc,BBFM17} and effective field theory (EFT) techniques \cite{FStail,FS4PN,FMSS16,FS19,FPRS19,Blumlein20}. The dissipative piece is known at 4.5PN~\cite{BuTh70, IW93, IW95, PW02, KFS03, NB05,itoh3, BFT24}~(see also  \cite{GII97,leibovich:2023xpg} for incomplete 4.5PN results).  Crucially, at 4PN, the equations of motion  feature a hereditary tail term, which depends on the whole history of the binary, rather than the instantaneous values of the positions and velocities. This tail term contributes both in the conservative and dissipative sectors. The other, instantaneous contributions neatly split into even and odd PN orders: instantaneous conservative contributions enter at 0PN, 1PN, 2PN, 3PN, and 4PN, whereas instantaneous dissipative contributions enter at 2.5PN, 3.5PN, and 4.5PN.

When considering only the \textit{conservative} dynamics, we expect the existence of \textit{conserved}, Noetherian quantities. These are the ten Poincaré invariants of Minkowski spacetime: the energy $E_\text{cons}$ (associated to time translations), the angular momentum $\mathbf{J}_\text{cons}$ (associated to spatial rotations), the linear momentum $\mathbf{P}_\text{cons}$ (associated to spatial translations) and the center-of-mass position $\mathbf{G}_\text{cons}$ (associated to boosts). These quantities are constant in time, so long as any acceleration appearing when taking a time-derivative is replaced by its conservative expression only.  

Once we know the equations of motion, it is then possible to compute the gravitational waveform, and in particular the fluxes associated to the Poincaré invariants, the most important one for data analysis being the energy flux. Currently, the energy flux is known at 4.5PN order~\cite{BFHLTa,BFHLTb} using the post-Newtonian multipolar post-Minkowskian (PN-MPM) framework. It was also obtained at 3PN using EFT techniques \cite{Amalberti:2024jaa} and at 2PN using the direct integration of the relaxed field equations (DIRE)~\cite{WW96}.

The quantity which is best measured by gravitational wave detectors is the phase of the $(2,2)$ mode of the gravitational wave, which is related to the orbital phase through known relations which depend on the choice of spacetime foliation. In standard PN counting conventions, the phasing starts at $-2.5$PN, and is monochromatic, i.e., linear in time ($\phi(t) = \omega_0t$). The first deviation from a monochromatic phasing (due to radiation reaction) enter at Newtonian order. Strictly speaking, the previous results give access only to the phase at 2PN order.  
However, we can invoke flux-balance arguments to improve our phase prediction by 2.5PN orders, namely up to 4.5PN order. For this, we consider the flux balance law, 

\begin{equation}
\frac{\dd E}{\dd t} = -\mathcal{F}_E \,.
\end{equation}
where $\mathcal{F}_E$ is the energy flux measured at infinity and $E$ is the \textit{binding energy} of the system, also called the \textit{Bondi energy}~\cite{LBW12} (we will ignore black hole absorption effects until indicated explicitly). This binding energy is related to the total ADM mass of the spacetime~\cite{ADM}, denoted $\dM$, which is strictly conserved and reads\footnote{The flux at 4.5PN required the knowledge of the ADM mass at 3PN order: the second term in the right-hand side of \eqref{eq:ADM_mass_definition} was therefore included in \cite{BFHLTa, BFHLTb}, but the third term was neglected, since it is a 3.5PN correction $[\sim \mathcal{O}(c^{-7})]$ that will first enter the 5PN flux.}
\begin{equation}\label{eq:ADM_mass_definition}
    \dM = m + \frac{E(t)}{c^2} + \frac{1}{c^2}\int_{-\infty}^{t}\dd t' \mathcal{F}_E(t')\,.
\end{equation}
Moreover, the binding energy is  related to the conservative energy by 
\begin{equation}
    E = E_\text{cons}+E_\text{diss} \,,
\end{equation}
where $E_\text{diss}$ are dissipative correction due to radiation reaction, starting at 2.5PN order. In analogy to electromagnetism, these are commonly referred to as \textit{Schott~terms}~\cite{Schott}. In work to appear soon~\cite{GLTP25}, collaborators have shown how this relates to the Wald-Zoupas prescription for constructing conserved quantities in the presence of dissipation \cite{Wald:1999wa, Grant:2021sxk}.

In the case of nonspinning circular orbits, if $E$ and $\mathcal{F}_E$  are known in terms of the orbital frequency $\omega$, the chain rule yields
\begin{equation}
    \frac{\dd \omega}{\dd t} = - \frac{\mathcal{F}_E(\omega)}{\dd E/ \dd \omega}\,,
\end{equation}
The flux and energy (including both conservative and Schott terms) being known at 4.5PN order, this equation gives us access to the 4.5PN phasing (compare to the 2PN phasing obtained without invoking the flux balance law).

Standard lore \cite{breuer_radiation_1981} goes that the Schott terms approximately vanish for quasicircular orbits\footnote{This is also true for orbit-averaged quasielliptic orbits.}, because they can be expressed as total derivatives and are interpreted as reversible exchanges of energy between the binary system and the gravitational field. Hence, the first nonvanishing contribution from the 2.5PN Schott term   for circular orbits in fact first enters at 5PN order. Thus, Schott terms were historically never considered for circular orbits. In this work, I will show that at 4PN, a new pseudo-Schott term arises due to the dissipative piece of the tail entering the equations of motion. I call it a pseudo-Schott term because it is not a total derivative, but rather is a (doubly) hereditary contribution, and most importantly, because it does not vanish for circular orbits. As we will see, the inclusion of this contribution is paramount for a fully self-consistent calculation of the 4.5PN phasing.

In Ref.~\cite{BFHLTa,BFHLTb}, collaborators and I obtained for the first time this 4.5PN phase, but using a slightly different approach. Building on many years of previous work within the group~\cite{MHLMFB20, MQ4PN_IR, MQ4PN_renorm, MQ4PN_jauge, Henry:2021cek, TLB22, TB23_ToM}, we obtained the 4.5PN flux in terms of the orbital frequency~\cite{BFHLTa,BFHLTb}, and realized that there was a residual dependence in an arbitrary parameter $b_0$, which originates from the coordinate transformation between the harmonic coordinates in the near zone and the radiative coordinates in the exterior vacuum zone~\cite{B87,TLB22}. This parameter can be interpreted as a choice in the foliation (or slicing) of spacetime. Then, we realized that the orbital frequency was not an observable at future null infinity, as it differed from the waveform frequency at 4PN order. We could then express the orbital frequency in terms of the waveform frequency associated to the $(\ell,m)=(2,2)$ mode, and the relation between these two frequencies involves the $b_0$ constant.
When expressing the flux in terms of the observable waveform frequency using this relation, we found that the $b_0$ dropped out, as expected physically. However, we were then left with a conundrum. On the one hand, we had the \textit{conservative} energy expressed in terms of the \textit{orbital} frequency, and this relation did not depend on $b_0$; on the other hand, we had the flux in terms of the \textit{waveform} frequency, which also did not depend on $b_0$. Since these were different frequencies, we seemingly could not relate them using the flux balance law without reexpressing one frequency in terms of the other --- but then, we would introduce an undesired $b_0$. Thus, we came up with a physical postulate, inspired by Detweiler~\cite{Detweiler:2005kq}, and described after Eq.~(6.13) of Ref.~\cite{BFHLTb}: if the compact binary is in fact a binary pulsar system, the emitted frequency of the electromagnetic pulses would be the orbital frequency, but this would be seen asymptotically as the waveform frequency, as it would be deformed by the spacetime curvature in the form of tails. Thus, we postulated that the \textit{binding} energy entering the flux balance law should have the same functional expression in terms of the \textit{waveform} frequency as the \textit{conservative} energy in terms of the \textit{orbital} frequency. In this work, I show rigorously that this postulate is indeed entirely correct at 4PN order. The result obtained using this postulate exactly agrees with the result found when consistently including the pseudo-Schott term at 4PN.

The rest of this work is organized as follows. In Sec.~\ref{sec:eom}, the reader is reminded of the structure of the tail term in the 4PN acceleration, and how it splits into conservative and dissipative contributions. In Sec.~\ref{sec:schott_general}, the general expressions of the 4PN pseudo-Schott terms are computed for all ten Poincaré invariants. In Sec.~\ref{sec:postadiabatic},  postadiabatic integration formulas, which were known for retarded time, are extended to the case of advanced time. In Sec.~\ref{sec:schott_circular}, the pseudo-Schott terms are computed for circular orbits. In Sec~\ref{sec:circular_binding_energy}, I show that the binding energy and angular momentum thus obtained are (i) independent of $b_0$ once they are expressed in terms of the waveform frequency, (ii) exactly equal to those obtained using the heuristic arguments stated previously, and (iii) are related by a first law of binary black hole mechanics (with the masses held constant), but in terms of the waveform frequency rather than the usual orbital frequency. In Sec.~\ref{sec:phasing}, I compute the orbital phasing and frequency evolution (including horizon effects), and relate it to the known waveform phasing \cite{BFHLTa,BFHLTb}. Finally, in Sec.~\ref{sec:discussion}, I discuss consequences of these results, as well as potential extensions of the physical postulate to higher PN order~\cite{GLTP25}.

\section{The 4PN equations of motion}
\label{sec:eom}
One of the main predictions of post-Newtonian techniques is how compact binaries move. This can be described using Lagrangians or Hamiltonians, but we will here prefer to give the explicit differential equations satisfied by the particle trajectories, which we call equations of motion. Namely, the acceleration of the first particle, denoted $\bm{a}_1 = \dd^2 \bm{y}_1/\dd t^2$, is given explicitly in terms of the positions and velocities of the two particles, $(\bm{y}_1,\bm{y}_2,\bm{v}_1,\bm{v}_2)$. The acceleration of the second particle, $\bm{a}_2$, can then be immediately deduced by swapping labels, $1\leftrightarrow 2$.
At 4PN, a new feature arises: the equations of motion do not only depend on the positions and velocities of the particles at time $t$, but become a functional of the positions and velocities of the particles throughout their past history, namely \mbox{$t' \mapsto \{\bm{y}_1(t'),\bm{y}_2(t'),\bm{v}_1(t'),\bm{v}_2(t')\}$} for all $t'<t$ . In other words, the differential equation governing the motion becomes an integro-differential equation. This gives rise to a wealth of difficulties, which are complemented by independent difficulties due to regularization that appear at 4PN as well.

These equations of motion have been successfully derived using three different techniques:  the Arnowitt–Deser–Misner (or Hamiltonian) formalism~\cite{BiniD13, DJS14,Damour:2014jta,JaraS15,DJS15eob, DJS16}, using the Fokker action \cite{BBBFMa,BBBFMb,BBBFMc,BBFM17} and using effective field theory techniques \cite{FStail,FS4PN,FMSS16,FS19,FPRS19,Blumlein20}.  Despite initial discrepancies, these three methods have converged toward a unique prediction for the conservative piece of the 4PN equations of motion. By conservative piece, I mean the time-even contributions to the equations of motion, which do not give rise to any energy loss of the binary system through gravitational radiation.

Restricting for now  our attention to the instantaneous piece of the equations of motion, one can identify the conservative sector to integer orders in the PN counting, namely the  Newtonian, 1PN, 2PN, 3PN, and 4PN contributions. Still restricting to the instantaneous piece, the dissipative contributions, which correspond to radiation reaction, arise at odd integers orders in the PN counting, namely  2.5PN, 3.5PN, and 4.5PN, and have been obtained  in Refs.~\cite{BuTh70, IW93, IW95, NB05, BFT24}. 

The hereditary tail contribution, which arises at 4PN order, exhibits in fact both a conservative and a dissipative contribution. These have been studied in the context of near-zone calculations a long time ago~\cite{BD88,BD92,B93,B97}, and more recently in the context of the full 4PN equations of motion~\cite{BiniD13, DJS14,Damour:2014jta,JaraS15,DJS15eob, DJS16, BBBFMa,BBBFMb,BBBFMc,BBFM17,FStail,FS4PN,FMSS16,FS19,FPRS19,Blumlein20}. 
To make things explicit, I will split the acceleration into an instantaneous and a tail part as follows:
\begin{align}
    a_1^i = a_{1,\text{inst}}^i + a_{1,\text{tail}}^i \,.
\end{align}
As discussed, the instantaneous part is well known, both in the conservative and the dissipative sector. I will focus on the hereditary tail, which simply reads~\cite{Damour:2014jta,BBFM17} 
\begin{align}\label{eq:a1_tail}
    a_{1,\text{tail}}^i &= - \frac{8 G^2 \dM}{5 c^8} y_1^j \Bigg\{\int_0^{+\infty} \dd\tau\, \ln\left(\frac{c\tau}{2 s_0}\right) \dM_{ij}^{(7)}(t-\tau) \nonumber\\*
    &\qquad\qquad\qquad - \dM_{ij}^{(6)} (t) \ln\left(\frac{r_{12}}{s_0}\right)\Bigg\}\,.
\end{align}
Here, $s_0$ is an arbitrary constant which I will discuss later; $\dM_{ij}$ is the canonical, mass-type, quadrupole moment of the system, and $\dM_{ij}^{(n)}$ is its $n$-th time derivative. Indeed, in the far zone, one performs a multipolar post-Minkowskian (MPM) iteration, see~\cite{BD88,Blanchet:2013haa} for details. The seed of this iteration is the linearized metric, which is parametrized in terms of canonical\footnote{One could have also considered the source quadrupole moment~$\mathrm{I}_{ij}$, which differs from the canonical moment at 2.5PN \cite{MQ4PN_jauge}. But for circular orbits, this 2.5PN term vanishes, such that the two are exactly equal, so we do not need to worry about such subtleties here.} moments $(\dM_L, \mathrm{S}_L)$. For our purposes, we only need to consider two types of moments: the monopole $\dM$, which corresponds to the constant ADM mass of the spacetime [see Eq.~\eqref{eq:ADM_mass_definition}], and the mass quadrupole $\dM_{ij}$.  In this work, only the leading order of the relation~\eqref{eq:ADM_mass_definition} is needed, which simplifies to $\dM = m + \mathcal{O}(1/c^2)$. The canonical quadrupole moment, $\dM_{ij}$, is obtained through asymptotic matching to the near zone, and is known up to 4PN order~\cite{MHLMFB20, MQ4PN_IR, MQ4PN_renorm,MQ4PN_jauge,BFHLTb}, and   it will only be required at 2.5PN in this work.

Importantly, this moment, which is a function of time, is required to fall off fast enough in the past to ensure convergence of the integral in \eqref{eq:a1_tail}. This is ensured in a binary system by the rate of the radiation-reaction-driven `outspiral' of the binary as one goes backwards in time. In practice however, one finds that the integral only depends on the very recent past \cite{BD88}. At leading order, the tail integral in fact does not depend on the rate of inspiral at all: this dependence only enters at relative~2.5PN through postadiabatic terms \cite{BFHLTb}, for which the convergence of the integral is not essential either.

Moreover, note that there is to some extent an arbitrary choice in how I split the hereditary part from the instantaneous part in Eq.~\eqref{eq:a1_tail}. As long as the tail piece contains all hereditary features, the choice is valid as long as one is consistent, and I have made a choice here which will be most convenient for our purposes. Consequently, in this expression, $s_0$~is an arbitrary, constant regularization scale (with the dimension of a length), which is allowed to differ from any other scale in the problem.  However, the dependence on $s_0$ in the hereditary term clearly cancels out against the instantaneous logarithmic term. We are therefore allowed to choose any scale which is convenient. In most previous works (e.g.~\cite{BBFM17}), the dependence in $s_0$ is explicitly eliminated in favor of $r_{12}(t)$. However, the time dependence in $r_{12}(t)$ can potentially generate unnecessary difficulties, in particular when evaluating the tails for circular orbits using postadiabatic integration formulas~\cite{BFHLTb} or in the presence of two nested time integrals~\cite{TB23_ToM}. Thus, to simplify expressions while avoiding these complications, we will choose from now on $s_0 = b_0$, where $b_0$ is the only other arbitrary scale remaining in the problem: it is linked to the relation between harmonic and radiative coordinates, which reads~\cite{B87,TLB22}
\begin{align}
    x_{\text{rad}}^{\mu} = x_{\text{harm}}^{\mu} + \frac{2 G \dM}{c^3} \,\delta^{0\mu} \ln\left( \frac{r_{\text{harm}}}{b_0}\!\right) + \mathcal{O}(G^2)\,,
\end{align}
where nontrivial corrections enter in the $\mathcal{O}(G^2)$ term.
Another way to interpret $b_0$ is as a choice of spacetime foliation or slicing: $b_0$ parametrizes a one-parameter family of foliations, which allow us to define simultaneous events in the near zone and at future null infinity, which is very useful in the context of flux balance laws. This choice of $s_0$ is thus entirely harmless and mostly aesthetic, as it avoids carrying around terms proportional to $\ln(s_0/b_0)$. 

As stated, the tail acceleration has both conservative and dissipative contributions. Formally, these can be untangled by dividing this integral in time-even and time-odd contributions. 
Under a time reflection $t\rightarrow -t$, the positions transform as $\bm{y}_{1,2} \rightarrow \bm{y}_{1,2}$ and the velocities  $\bm{v}_{1,2} \rightarrow -\bm{v}_{1,2}$. From the explicit expressions of the multipolar moments in terms of $(\bm{y}_{1,2},\bm{v}_{1,2})$, one  then finds that $\dM_{ij}^{(n)} \rightarrow (-)^n\dM_{ij}^{(n)}$, which implies that $\dM^{(n)}_{ij}(t\pm\tau) \rightarrow (-)^n \dM_{ij}^{(n)}(t\mp\tau)$. These symmetry considerations lead to the split~\cite{BBFM17,GLPR16}
\begin{widetext}
\begin{subequations}\label{eq:a1_tail_cons_diss}\begin{align}
\label{seq:a1_tail_cons}
    a_{1,\text{tail, cons}}^i &= - \frac{4 G^2 \dM}{5 c^8} y_1^j \Bigg\{ \int_0^{+\infty} \dd\tau\, \ln\left(\frac{c\tau}{2 b_0}\right) \left[\dM_{ij}^{(7)}(t-\tau) - \dM_{ij}^{(7)}(t+\tau) \right] - 2 \dM_{ij}^{(6)} \ln\left(\frac{r_{12}}{b_0}\right) \Bigg\} \,,\\*
\label{seq:a1_tail_diss}
     a_{1,\text{tail, diss}}^i &= - \frac{4 G^2 \dM}{5 c^8} y_1^j \int_0^{+\infty} \dd\tau\, \ln\left(\frac{c\tau}{2 b_0}\right) \left[\dM_{ij}^{(7)}(t-\tau) + \dM_{ij}^{(7)}(t+\tau) \right] \,, 
\end{align}\end{subequations}
\end{widetext}
where the conservative piece is manifestly invariant under the time reflection (i.e., time-even) and the dissipative piece flips sign (i.e., time-odd). 
Evidently, such a definition relies on the assumption that the quadrupole moment falls off fast enough not only in the past, but also in the future. The falloff in the future is in fact true for the fully relativistic problem, since the two black holes eventually merge and ring down exponentially. However, within the post-Newtonian framework, this is not the case, because for bound systems, these quadrupolar moments \textit{blow up in finite time}, due to the fact that the waveform amplitudes and frequencies become infinite at ``coalescence time'', usually denoted $t_c$. Since the ill-defined advanced piece was introduced artificially to separate conservative and dissipative contributions, we are entitled to regularize it as we wish, so long as the same regularization is used in both sectors. In practice, the fact that the tail integrals depend only on the recent past  \cite{BD88} (and now, near future)  will have no impact on our formulas.

\section{General expression of the Schott term}
\label{sec:schott_general}

From the Hamiltonian generating the conservative acceleration, it is possible to construct the ten (conservative) Poincaré invariants associated with the symmetries of Minkowski spacetime: the energy $E_\text{cons}$ (associated with time translations), the angular momentum $J^i_\text{cons}$ (associated with rotations), the linear momentum $P^i_\text{cons}$ (associated with spatial translations) and the center-of-mass position $G^i_\text{cons}$ (associated with boosts). These invariants, which we denote collectively as $\bm Q \equiv (E, J^i, P^i, G^i)$,  are expressed in terms of $(\bm{y}_1,\bm{y}_2,\bm{v}_1,\bm{v}_2)$ and are constant when taking time derivatives using the purely conservative equations of motion, namely

\begin{align}
    \left.\frac{\dd \bm{Q}_{\text{cons}}}{\dd t}\right|_{\bm{a}_{1}^{\text{cons}}} \!\!\! = 0 \, .
\end{align}

Note that the conservative energy is obtained at 3PN as the on-shell value of the Hamiltonian, but this is no longer the case at 4PN due to the presence of hereditary tails in the Hamiltonian~\cite{BBFM17,Blanchet:2017rcn} (see also \cite{DJS15eob, DJS16} for an alternative approach using Delaunay averaging).   Moreover, the 4PN energy does not only depend on the instantaneous values of the phase variables anymore, but also has a functional dependence on the whole history of the binary. Similar observations apply to the other Poincaré invariants.

In parallel to this near-zone computation, one can compute the various fluxes at future null infinity associated to these invariants. These are known in terms of the radiative  moments $(\dU_L, \dV_L)$, which describe the asymptotic waveform, and read~\cite{Thorne:1980ru, Blanchet:2013haa, BF18, COS19}
\begin{subequations}\label{eq:fluxes_UL_VL}\begin{align}
\label{seq:flux_E_UL_VL}\mathcal{F}_E &= \sum_{\ell=2}^\infty \frac{G}{c^{2\ell + 1}} \bigg\{a_\ell \dU_L^{(1)}\dU_L^{(1)}+ \frac{b_\ell}{c^2} \dV_L^{(1)}\dV_L^{(1)}\bigg\} \,,\\
\label{seq:flux_J_UL_VL}
    \mathcal{F}^i_{\bm{J}} &= \epsilon_{ijk}\sum_{\ell=2}^\infty \frac{G}{c^{2\ell + 1}} \bigg\{a'_\ell \dU_{jL-1}\dU_{kL-1}^{(1)}\nonumber\\*
    & \qquad\qquad\qquad\quad+ \frac{b'_\ell}{c^2} \dV_{jL-1} \dV_{kL-1}^{(1)}\bigg\} \,,\\
\label{seq:flux_P_UL_VL}
    \mathcal{F}^i_{\bm{P}} &= \sum_{\ell=2}^\infty \frac{G}{c^{2\ell + 3}} \bigg\{a''_\ell \dU_{iL}^{(1)}\dU_L^{(1)}+ \frac{b''_\ell}{c^2} \dV_{iL}^{(1)}\dV_L^{(1)} \nonumber\\*
    & \qquad\qquad\qquad+ c''_\ell \epsilon_{ijk} \dU_{jL-1}^{(1)} \dV_{kL-1}^{(1)}\bigg\} \,,\\
\label{seq:flux_G_UL_VL}
    \mathcal{F}^i_{\bm{G}} &= \frac{1}{2} \sum_{\ell=2}^\infty \frac{G}{c^{2\ell + 3}} \bigg\{a'''_\ell \left(\dU_{iL} \dU_L^{(1)} -\dU_{iL}^{(1)}\dU_L \right) \nonumber\\*
    & \qquad\qquad\qquad + \frac{b'''_\ell}{c^2} \left(\dV_{iL}\dV_L^{(1)}-\dV_{iL}^{(1)}\dV_L\right) \bigg\} \,,
\end{align}
\end{subequations}
where the coefficients read \cite{Blanchet:2013haa}
\begin{align}
    a_\ell &=\frac{(\ell \!+\! 1)(\ell \!+\! 2)}{(\ell \!-\! 1) \ell \ell! (2\ell \!+\! 1)!!} \,, & b_\ell &= \frac{4\ell(\ell \!+\! 2)}{(\ell \!-\! 1)(\ell \!+\! 1)!(2\ell \!+\! 1)!!} \,,\nonumber\\
    a'_\ell &= \frac{(\ell \!+\! 1)(\ell \!+\! 2)}{(\ell \!-\! 1)  \ell! (2\ell \!+\! 1)!!}\,, & b'_\ell &= \frac{4\ell^2(\ell \!+\! 2)}{(\ell \!-\! 1)(\ell \!+\! 1)!(2\ell \!+\! 1)!!}\,,\nonumber\\
    a''_\ell &= \frac{2(\ell \!+\! 2)(\ell \!+\! 3)}{\ell (\ell \!+\! 1)! (2\ell \!+\! 3)!!}\,, & b''_\ell &= \frac{8 (\ell \!+\! 3)}{(\ell \!+\! 1)!(2\ell \!+\! 3)!!}\,,\nonumber\\
    a'''_\ell &= \frac{2(\ell \!+\! 2)(\ell \!+\! 3)}{\ell \ell! (2\ell \!+\! 3)!!}\,, & b'''_\ell &= \frac{8 (\ell \!+\! 3)}{\ell!(2\ell \!+\! 3)!!}\,, \nonumber 
\end{align}
\vspace{-0.75cm}
\begin{align}
c''_\ell = \frac{8(\ell \!+\! 2)}{(\ell \!-\! 1)(\ell \!+\! 1)!(2\ell \!+\! 1)!!} \,.\hspace{3.25cm}
\end{align}
The fluxes associated to energy and angular momentum  enter at 2.5PN, whereas the fluxes associated to linear momentum and center-of-mass position enter at 3.5PN order. Moreover, note that here I have adopted a definition for the flux associated to the center-of-mass position borrowed in Ref.~\cite{COS19}, which stems from a computation purely at future null infinity, rather than the definition given in Refs.~\cite{BF18, BFT24}, which stems from a near-zone computation  (these two definitions differ only by a total time derivative). Typically, the MPM algorithm allows us to express the radiative moments in terms of, e.g., canonical moments $(\dM_L, \mathrm{S}_L)$, which can themselves be expressed in terms of the post-Newtonian source through matching. For example, the mass-type quadrupole moments are related at 1.5PN order by
\begin{align}
    \dU_{ij} &= \dM_{ij}^{(2)}\! + \frac{2G \dM}{c^3}\!\int_0^{+\infty} \!\!\!\! \dd \tau \! \left[\ln\left(\frac{c\tau}{2b_0}\right) + \frac{11}{12}\right]\! \dM_{ij}^{(4)}(t-\tau) \,,
\end{align}
We now want to establish flux balance laws, which relate quantities computed in the near-zone to quantities computed at future null infinity. To this end,  the full acceleration is reinstated, which contains both conservative and dissipative contributions: \mbox{$\bm{a}_{1}= \bm{a}_{1}^{\text{cons}} +\bm{a}_{1}^{\text{diss}}$}. This leads to the general structure 
\begin{align}\label{eq:Schott_general_equation_1}
    \left.\frac{\dd \bm{Q}_\text{cons}}{\dd t}\right|_{\bm{a}_1} + \mathcal{F}_{\bm{Q}} = - \delta \mathcal{F}_{\bm{Q}}\,,
\end{align}
where the right-hand side is a small, dominantly 2.5PN quantity for $\left\{E, \bm{J}\right\}$, and an even smaller 3.5PN quantity for $\left\{ \bm{P}, \bm{G}\right\}$.  When ignoring 4PN contributions, it was shown in \cite{IW93,IW95, BFT24} that this term is in fact a total time-derivative, such that $\delta \mathcal{F}_{\bm{Q}} = {\dd \bm{Q}_\text{diss}}/{\dd t}$, where the radiation-reaction contribution $\bm{Q}_\text{diss}$ is called a \textit{Schott term}, in reference to similar terms in electromagnetism \cite{Schott}. One then defines the \textit{binding} quantities for the full dissipative system as $\bm{Q} = \bm{Q}_\text{cons} + \bm{Q}_\text{diss}$, such that the expected flux balance law be satisfied
\begin{align}
    \left.\frac{\dd \bm{Q}}{\dd t}\right|_{\bm{a}_1} + \mathcal{F}_{\bm{Q}} = 0 \,.
\end{align}
Moreover, general arguments~\cite{breuer_radiation_1981} predict that these Schott terms themselves can in fact be written as total derivatives (again, when ignoring tails).  For example, at 2.5PN order, we have~\cite{BFT24} 
\begin{align}
    E^{\text{2.5PN}}_{\text{diss}} &= \frac{G}{5 c^5}\Big\{\dM_{ij}^{(2)}\dM_{ij}^{(3)}-\dM_{ij}^{(1)}\dM_{ij}^{(4)}\Big\}\nonumber\\*
    &= \frac{G}{5 c^5} \frac{\dd}{\dd t}\Bigg[\dM_{ij}^{(2)}\dM_{ij}^{(2)} - \dM_{ij}^{(1)}\dM_{ij}^{(3)}\Bigg] \,,
\end{align} 
although checking this statement explicitly at higher orders would require more work, along the lines of~Ref.~\cite{BFT24}.
A consequence is that these Schott terms vanish after orbit averaging for bound elliptic systems, or identically for circular orbits. This can also be seen by the fact that the explicit expressions of the Schott terms given in~\cite{BFT24} are all proportional to $\dot{r}$, which vanishes for exactly circular orbits. Thus, the Schott term  is  usually interpreted as a reversible exchange of energy between the binary and the gravity field.  However, due to radiation reaction, orbits are not perfectly circular: the slow inspiral causes the 2.5PN Schott term to start contributing, in the case of circular orbits, at 5PN order~\cite{breuer_radiation_1981}. 

The results obtained in Ref.~\cite{BFT24} purposely ignore the known dissipative contributions that enter at 4PN order, due to tails. A detailed treatment of the associated Schott terms is the main object of this work. In order to work these out,  we will only require the Newtonian expressions of the Poincaré invariants
\begin{subequations}\label{eq:newtonian_poincare_invariants}\begin{align}
    E_\text{cons} &= \frac{1}{2}\left(m_1 v_1^2 + m_2 v_2^2\right) + \mathcal{O}\left(\frac{1}{c^2}\right) \,,\label{seq:newtonian_E}\\
    J^i_\text{cons} &= \epsilon_{ijk} \left(m_1 y_1^j v_1^k+m_2 y_2^j v_2^k\right) + \mathcal{O}\left(\frac{1}{c^2}\right)\,, \label{seq:newtonian_J}\\
     P^i_\text{cons} &=  m_1 v_1^i + m_2 v_2^i  + \mathcal{O}\left(\frac{1}{c^2}\right) \,,\label{seq:newtonian_P}\\
     G^i_\text{cons} &=  m_1 y_1^i + m_2 y_2^i  + \mathcal{O}\left(\frac{1}{c^2}\right) \,.\label{seq:newtonian_G}
\end{align}\end{subequations}
Take the time derivatives of these quantities, and discard all contributions which do not involve the 4PN dissipative acceleration~\eqref{seq:a1_tail_diss}, which is only due to tails. Recognizing at this order \mbox{$2(m_1 v_1^{\langle i} y_1^{j \rangle} + m_2 v_2^{\langle i} y_2^{j \rangle}) = \dM_{ij}^{(1)}$}, \mbox{$m_1 y_1^{\langle i} y_1^{j \rangle} + m_2 y_2^{\langle i} y_2^{j \rangle} = \dM_{ij}$}, and \mbox{$m_1 y_1^i + m_2 y_2^i = \dM_i$}, one finds expressions that agree, up to the arbitrary choice of scale, with~(6.11)~of~\cite{BBFM17}
\begin{widetext}
\begin{subequations}\label{eq:dHdt_diss}\begin{align}\label{seq:dEdt_diss}
     \left.\frac{\dd E_{\text{cons}}}{\dd t}\right|_{\bm{a}_{1,\text{diss}}^{\text{tail}}} &= - \frac{2 G^2 \dM}{5 c^8} \dM_{ij}^{(1)} \int_0^\infty \dd \tau \ln\left(\frac{c\tau}{2 b_0}\right) \left[\dM_{ij}^{(7)}(t-\tau) + \dM_{ij}^{(7)}(t+\tau) \right] + \mathcal{O}\left(\frac{1}{c^{10}}\right) \,, \\*
\label{seq:dJidt_diss}
     \left.\frac{\dd J^{i}_{\text{cons}}}{\dd t}\right|_{\bm{a}_{1,\text{diss}}^{\text{tail}}} &= - \frac{4 G^2 \dM}{5 c^8} \epsilon_{ijk }\dM_{jl}  \int_0^\infty \dd \tau \ln\left(\frac{c\tau}{2 b_0}\right) \left[\dM_{kl}^{(7)}(t-\tau) + \dM_{kl}^{(7)}(t+\tau) \right] + \mathcal{O}\left(\frac{1}{c^{10}}\right)\,, \\*
\label{seq:dPidt_diss}
      \left.\frac{\dd P^{i}_{\text{cons}}}{\dd t}\right|_{\bm{a}_{1,\text{diss}}^{\text{tail}}} &= - \frac{4 G^2 \dM}{5 c^8} \dM_{j} \int_0^\infty \dd \tau \ln\left(\frac{c\tau}{2 b_0}\right) \left[\dM_{ij}^{(7)}(t-\tau) + \dM_{ij}^{(7)}(t+\tau) \right] + \mathcal{O}\left(\frac{1}{c^{10}}\right)\,, \\*
\label{seq:dGidt_diss}
      \left.\frac{\dd G^{i}_{\text{cons}}}{\dd t}\right|_{\bm{a}_{1,\text{diss}}^{\text{tail}}} &=  \mathcal{O}\left(\frac{1}{c^{10}}\right)
\end{align}\end{subequations}
\end{widetext}
The right-hand side of Eq.~\eqref{seq:dGidt_diss} vanishes because taking one time-derivative of Eq.~\eqref{seq:newtonian_G} does not generate any accelerations. Moreover, note that in the center-of-mass frame, one has $\dM_i = \mathcal{O}\left(1/c^7\right)$, and the right-hand side of~\eqref{seq:dPidt_diss} vanishes. 

We now focus on what goes on at future null infinity. We know that the 1.5PN contribution to the fluxes at infinity are entirely due to tails, and read
\begin{widetext}
\begin{subequations}\begin{align}
    \mathcal{F}^{\text{tail}}_E &= \frac{4 G^2 \dM}{5 c^8} \dM_{ij}^{(3)}\int_0^{+\infty} \! \dd\tau \left[\ln\left(\frac{c\tau}{2 b_0}\right)  + \frac{11}{12} \right]\dM_{ij}^{(5)}(t-\tau) \,,\\*
    \mathcal{F}^{i,\text{tail}}_{\bm{J}}  &= \frac{4 G^2 \dM}{5 c^8}\epsilon_{ijk}\Bigg\{\dM_{jl}^{(2)}\int_0^{+\infty}  \!\dd\tau   \left[\ln\left(\frac{c\tau}{2 b_0}\right) + \frac{11}{12} \right] \dM_{kl}^{(5)}(t-\tau) \nonumber\\*
    &\qquad\qquad\ \quad - \dM_{jl}^{(3)}\int_0^{+\infty} \! \dd\tau   \left[\ln\left(\frac{c\tau}{2 b_0}\right) + \frac{11}{12} \right] \dM_{kl}^{(4)}(t-\tau)\Bigg\} \,,\\*    
    \mathcal{F}^{i,\text{tail}}_{\bm{P}} &=0\,,\\*
    \mathcal{F}^{i,\text{tail}}_{\bm{G}} &=0\,.
\end{align}\end{subequations}
\end{widetext}
The vanishing of the fluxes associated to the linear momentum and the center-of-mass position at this order is due to the conservation of the ADM linear momentum (i.e, the canonical  dipole moment).
Thus, it is clear that the flux balance law for the energy, angular momentum and linear momentum are not automatically satisfied, and indeed require the inclusion of a 4PN Schott term. Conversely, no such Schott term is required for the flux associated to the center-of-mass position at this order.

At this order, the tail sector can be treated independently from the instantaneous sector. Note that the conservative tails entering both the acceleration and the conservative Poincaré invariants are already accounted for and do not need to be treated again. Thus, we only need to focus on the contribution of the dissipative tails, and Eq.~\eqref{eq:Schott_general_equation_1} then immediately translates to
\begin{align}
\left.\frac{\dd \mathbf{Q}_\text{cons}^\text{N}}{\dd t}\right|_{\mathbf{a}_{1,\text{diss}}^\text{tail}} + \mathcal{F}_\mathbf{Q}^\text{tail} = - \delta \mathcal{F}_\mathbf{Q}^\text{tail} \,.
\end{align}
The associated Schott term is then given by
\begin{align}
\mathbf{Q}_\text{diss}^\text{4PN} = \int_{-\infty}^t \dd t' \,\delta    \mathcal{F}_\mathbf{Q}^\text{tail}(t')\,.
\end{align}
With a few operations by parts, we find that the Schott terms associated to the energy and angular momentum can be written as the total time derivative of some instantaneous and tail-like terms, plus a tail-like term which cannot be written as a total time derivative. Thus, the sought-for Schott term will necessarily feature a doubly hereditary term, akin the ``genuine'' tails of memory, see the first term with a double integral in Eq.~(6.6c)~of~\cite{TB23_ToM}. Note that the Schott term entering the linear momentum has a simpler structure, thanks to the conservation of both the ADM mass (monopole) and linear momentum (dipole).
Our final expressions for the Schott terms reads
\begin{widetext}
\begin{subequations}\label{eq:H_diss_4PN}\begin{align}\label{seq:E_diss_4PN}
      E^{\text{4PN}}_{\text{diss}}  &=  \frac{2 G^2 \dM}{5 c^8} \Bigg\{
     \int_0^\infty \dd \rho \, \dM_{ij}^{(4)}(t-\rho) \int_{0}^\infty \dd \tau \ln\left(\frac{c\tau}{2 b_0}\right) \left[\dM_{ij}^{(4)}(t-\rho-\tau) - \dM_{ij}^{(4)}(t-\rho+\tau)\right]\nonumber\\*
     &\qquad\qquad -\dM_{ij}^{(3)} \int_{0}^\infty  \dd\tau \ln\left(\frac{c\tau}{2 b_0}\right) \left[\dM_{ij}^{(4)}(t-\tau) -  \dM_{ij}^{(4)}(t+\tau) \right] \nonumber\\*
    &\qquad\qquad\ -\dM_{ij}^{(2)} \int_{0}^\infty  \dd\tau \ln\left(\frac{c\tau}{2 b_0}\right) \left[\dM_{ij}^{(5)}(t-\tau) +
    \dM_{ij}^{(5)}(t+\tau) \right] \nonumber\\*
    &\qquad\qquad\ +\dM_{ij}^{(1)} \int_{0}^\infty  \dd\tau \ln\left(\frac{c\tau}{2 b_0}\right) \left[\dM_{ij}^{(6)}(t-\tau) +
    \dM_{ij}^{(6)}(t+\tau) \right] \nonumber\\*
    &\qquad\qquad - \frac{11}{12} \dM_{ij}^{(3)} \dM_{ij}^{(3)}    
      \Bigg\} \,, \\
\label{seq:J_diss_4PN}
    J^{i,\text{4PN}}_{\text{diss}}  &=  \frac{4 G^2 \dM}{5 c^8} \epsilon_{ijk} \Bigg\{
     \int_0^\infty \dd \rho \, \dM_{jl}^{(3)}(t-\rho) \int_{0}^\infty \dd \tau \ln\left(\frac{c\tau}{2 b_0}\right) \left[\dM_{kl}^{(4)}(t-\rho-\tau) - \dM_{kl}^{(4)}(t-\rho+\tau)\right]\nonumber\\*
     &\qquad\qquad +\dM_{jl}^{(2)} \int_{0}^\infty  \dd\tau \ln\left(\frac{c\tau}{2 b_0}\right) \dM_{kl}^{(4)}(t+\tau)  \nonumber\\*
     &\qquad\qquad\ -\dM_{jl}^{(1)} \int_{0}^\infty  \dd\tau \ln\left(\frac{c\tau}{2 b_0}\right) \left[\dM_{kl}^{(5)}(t-\tau) +
    \dM_{kl}^{(5)}(t+\tau) \right] \nonumber\\*
    &\qquad\qquad\ +\dM_{jl} \int_{0}^\infty  \dd\tau \ln\left(\frac{c\tau}{2 b_0}\right) \left[\dM_{kl}^{(6)}(t-\tau) +
    \dM_{kl}^{(6)}(t+\tau) \right] \nonumber\\*
    &\qquad\qquad - \frac{11}{12} \dM_{jl}^{(2)} \dM_{kl}^{(3)}    
      \Bigg\} \,,\\
\label{seq:P_diss_4PN}
     P^{i,\text{4PN}}_{\text{diss}}  &= \frac{4 G^2 \dM}{5 c^8} \dM_{j} \int_0^\infty \dd \tau \ln\left(\frac{c\tau}{2 b_0}\right) \left[\dM_{ij}^{(6)}(t-\tau) + \dM_{ij}^{(6)}(t+\tau) \right] + \mathcal{O}\left(\frac{1}{c^{10}}\right)\,.
\end{align}\end{subequations}
\end{widetext}
Unlike the other Schott terms at 2.5PN, 3.5PN, and 4.5PN \cite{BFT24}, the expressions obtained in~\eqref{eq:H_diss_4PN}  are far from being total derivatives. They are not even an instantaneous function of the phase space variables due to the presence of tail terms, and the energy and angular momentum even feature a doubly hereditary integral. Thus, we expect these \textit{pseudo}-Schott terms to be nonvanishing in the circular limit. A tentative physical interpretation goes as follows: due to the presence of hereditary terms, this expression is sensitive to small, radiation-reaction induced variations in the amount of energy reversibly exchanged during one orbit. 

Finally, note that there are of course many other formulations of this Schott term, which can be obtained through integration by parts. Some of these formulations are simpler, others distinguish $s_0$ and $b_0$. The motivation for the formulation in Eq.~\eqref{eq:H_diss_4PN} is aesthetic: (i) there is only one doubly hereditary term; (ii) it is expressed in a canonical form (e.g., four time derivatives on each moment for the energy~\cite{TB23_ToM}); and (iii) the only arbitrary constant is $b_0$. I have checked that various such formulations yield the same explicit result when reduced to the case of circular orbits.

In association with the results of Refs.~\cite{BBFM17, BBFM17, BFT24}, these results allow us to completely control all ten Poincaré invariants at 4.5PN order, including both conservative and dissipative (Schott) contributions. Following the same argument as in the final paragraph of Sec. IV of Ref.~\cite{BBFM17}, we can see that Schott term in the linear momentum will  \textit{not} affect the 4.5PN-accurate formula used to express $(\bm{y}_1, \bm{y}_2,\bm{v}_1,\bm{v}_2)$ in terms of the center-of-mass variables $(\bm{x}, \bm{v})$, see Eq.~(6.5)~of~\cite{BFT24}. This is because the dipole moment vanishes in the center-of-mass frame at this order, namely $\dM_i = \mathcal{O}\left(1/c^7\right)$.

\section{Integration formulas for circular orbits}
\label{sec:postadiabatic}
\subsection{Postadiabatic integration formulas for advanced tails}
\label{subsec:postadiabatic_derivation}

\interfootnotelinepenalty=10000

 The explicit computation of the contribution of tails for quasicircular orbits was extensively studied. It was shown that these integrals are mostly sensitive to the recent past such that we can compute these integrals with a perfectly circular model up to 2PN relative order, arguing that the cumulative oscillatory features in the remote past average out~\cite{BS93}. It was later confirmed by a computation that accounted for the rate of  inspiral~\cite{ABIQ04}, showing that the integral was indeed convergent in the infinite past and that the result did not depend on the inspiral rate. Recently, this computation was extended to 2.5PN \cite{BFHLTb,Trestini:2023khz}, where one needs to account for postadiabatic effects. We introduce the orbital frequency $\omega$, the dimensionless PN counting parameter
 \begin{align}\label{eq:y_def}
    y &= \left(\frac{G m \omega}{c^3}\right)^{2/3}
 \end{align}
 and the adiabatic parameter\footnote{\label{footnote:xi}By definition, the system is adiabatic when $\xi \ll 1$, which here is equivalent to the post-Newtonian assumption $y\ll 1$. The $8/3$ prefactor is conventional~\cite{ABIQ04, Trestini:2023khz}, but was dropped in Ref.~\cite{BFHLTb}.} defined as  $\xi = 8 \dot{\omega}/(3\omega^2)$, and we find that the convergence is still ensured, and the rate of inspiral now enters explicitly in the result. The result is given in Eq.~(5.16) of \cite{BFHLTb}, and reads (after rescaling $\xi$; see Footnote~\ref{footnote:xi})
\begin{widetext}
    \begin{align}\label{eq:postadiabatic_retarded_convergent}
     &\int_0^{+\infty} \dd \tau \ln\left(\frac{\tau}{\tau_0}\right)  \left[y(t -\tau)\right]^\alpha \de^{- \di n \phi(t- \tau)} =  \frac{y^{\alpha}\de^{-\di n \phi} }{\di n \omega} \Bigg[ \left(1+\frac{2\alpha-3}{8} \, \frac{\xi}{\di n}\right)\left(\ln(|n| \omega \tau_0   ) + \gamma_\text{E} - \text{sg}(n) \frac{\di \pi}{2}\right) - \frac{4\alpha-9}{16} \frac{\xi}{\di n}\Bigg]\,,
 \end{align}
 \end{widetext}
where quantities are evaluated at time $t$ unless otherwise indicated.

\interfootnotelinepenalty=500
 
This computation, however, was done only for the physically relevant case where one integrates over the whole \textit{past} of the binary. When splitting our problem into conservative and dissipative pieces, we artificially created the need to compute tails spanning the whole \textit{future} of the binary. In the real, nonperturbative problem, this is expected to be entirely regular, since the system rings down after merger. However, in the perturbative problem we are considering, most physical quantities (e.g., the orbital frequency) blow up in finite time, and the blowup time can be interpreted as ``coalescence time'', denoted~$t_c$. The function we are trying to integrate around is not even defined after that time, and even if we change the upper bound to merger time, the integral doesn't converge. Since this ill-defined contribution was artificially introduced, it will suffice to find a prescription, so long as the same one is used in the conservative and dissipative sectors. One such prescription would be to introduce a regulator in the integral, e.g., $\de^{-\epsilon \, \xi(t+\tau)}$, which imposes a smoothly vanishing integrand at merger time for any $\epsilon>0$, since $\xi(t) \rightarrow + \infty$ at merger time ($t \rightarrow t_c$). One can then reproduce, \textit{mutatis mutandis}, the steps of Ref.~\cite{Trestini:2023khz}, and send $\epsilon \rightarrow 0$ at the end of the computation. However, such an approach was not successful in obtaining a clear derivation\footnote{More precisely, it was unclear how to compute terms of the type $\int_0^1 \dd s   \ln(s)\,\de^{-{8\di n s}/{(5 \xi)}} \, \de^{- \epsilon/(1-s)}$ without introducing terms involving $\mathrm{Ei}(-8\di n/(5\xi))$, which are undesired.}, although it was instructive in gaining intuition about how the postadiabatic formula should look in the advanced case.

A better strategy is  to \textit{define} the advanced case from the known retarded case using the time reflection symmetry discussed in Sec.~\ref{sec:eom}. Indeed, in the case of circular orbits, a time reflection $t\rightarrow -t$ leads to the following transformations: $\phi \rightarrow -\phi$, $\omega \rightarrow \omega$ and $\dot{\omega}\rightarrow - \dot{\omega}$. Automatically, we also have $y \rightarrow y$ and $\xi \rightarrow- \xi$. In particular, this means that $\phi(t\pm\tau) \rightarrow - \phi(t \mp \tau)$,  $\omega(t\pm\tau) \rightarrow \omega(t \mp \tau)$, $y(t\pm\tau) \rightarrow y(t \mp \tau)$ and $\xi(t\pm\tau) \rightarrow - \xi(t \mp \tau)$. 
Since we have established that the tails depend, in a loose sense, only on the recent past, we will borrow the notation of Eq.~(4.22)~of~Ref.~\cite{ABIQ04} and define
\begin{align}\label{eq:postadiabatic_def}
    \mathcal{I}_{\alpha,n}^{\pm} \equiv \int_0 \dd \tau \ln\left(\frac{\tau}{\tau_0}\right) \, \left[y(t \pm\tau)\right]^\alpha \de^{\pm\di n \phi(t\pm \tau)} \,.
\end{align}
Note that this is merely a convenient notation: an antiderivative is only defined up to an integration constant, which cannot be fixed with only one bound. We have ensured that the definition in the retarded case coincides with Eq.~\eqref{eq:postadiabatic_retarded_convergent}, whereas the advanced ($+$) case was defined from the retarded case ($-$) via time reflection symmetry. Now, applying the same time reflection on the right-hand side of Eq.~\eqref{eq:postadiabatic_def}, we find our postadiabatic formula for both the retarded and advanced cases must read
\begin{widetext}
    \begin{align}\label{eq:postadiabatic_integration}
    \mathcal{I}_{\alpha,n}^{\pm}  =  \de^{\pm\di n\phi}\frac{y^\alpha}{\di n\omega} \Bigg[\left(1\mp \frac{2\alpha-3}{8} \frac{\xi}{in}\right)\bigg(\ln\left(|n|\omega \tau_0\right) + \gamma_{\text{E}}  -  \,\mathrm{sg}(n) \frac{\di\pi}{2} \bigg)\pm \frac{4\alpha-9}{16} \frac{\xi}{\di n}\Bigg] + \mathcal{O}(\xi^2)\,.
\end{align}
\end{widetext}
This result is consistent with what was obtained in the incomplete approach using an $\epsilon$ regulator. 
At leading PN order, $\phi(t\pm\tau) \approx  \phi(t) \pm \tau \omega(t) +\cdots$, and the time evolution of $y$ can be neglected, so Eq.~\eqref{eq:postadiabatic_integration} reduces to the usual adiabatic formula
\begin{align}\label{eq:adiabatic_integration}
    &\int_{0} \dd \tau \,\ln\left(\frac{\tau}{\tau_0}\right)   \de^{\di n \omega \tau}  \nonumber\\*
    &= \frac{1}{\di n\omega} \Bigg[\ln\left(|n|\omega\tau_0\right) + \gamma_{\text{E}}  - \,\mathrm{sg}(n) \frac{\di \pi}{2} \Bigg] + \mathcal{O}(\xi)\,,
\end{align}
for which the influence of the frequency evolution is negligible. In fact, if one artificially imposes a nonradiative spacetime in which orbits are exactly circular, then one should set $\xi = 8\dot{\omega}/(3\omega^2) = 0$ in Eq.~\eqref{eq:postadiabatic_integration}, and one recovers the usual formula for the conservative tail, which would be valid in that scenario at 2.5PN order. However, note that in the physical case of a radiative spacetime, the time-even projection $\frac{1}{2}(\mathcal{I}_{\alpha,n}^{+}+\mathcal{I}_{\alpha,n}^{-})$ does depend on~$\xi$. The latter dependence of course only enters at 2.5PN order, such that the formula introduced here is consistent with the adiabatic formula used for the conservative dynamics \cite{Damour:2014jta, BBFM17}. This agreement is paramount for internal consistency of the formalism. 

\subsection{Self-consistency check for the postadiabatic tail formula using integration by parts}
\label{subsec:postadiabatic_test}

The previous derivation of the integration formula for the advanced tails was mostly based on symmetry considerations, and admittedly lacks a clear proof featuring convergent integrals. However, in this section, I will show that the formula derived  in Eq.~\eqref{eq:postadiabatic_integration} is the only one that is consistent with integration by parts at the level of the multipolar moments.

For definiteness, let us focus on the relation
\begin{align}\label{eq:IBP_test}
    \frac{\dd}{\dd t}&\Bigg[\dM^{(3)}_{ij} \int_0^{\infty}\dd\tau \ln\left(\frac{c \tau}{2 b_0}\right) \dM^{(4)}_{ij}(t\pm\tau)\Bigg] \nonumber\\*
    &= \dM^{(4)}_{ij} \int_0^{\infty}\dd\tau \ln\left(\frac{c \tau}{2 b_0}\right) \dM^{(4)}_{ij}(t\pm\tau) \nonumber\\*
    &\ +\dM^{(3)}_{ij} \int_0^{\infty}\dd\tau \ln\left(\frac{c \tau}{2 b_0}\right) \dM^{(5)}_{ij}(t\pm\tau) \,.
\end{align}
This relation is of course always true, and we will want to check that it still holds when replacing the multipolar moments by their explicit expressions for circular orbits. As we will see,  the postadiabatic formula~\eqref{eq:postadiabatic_integration} will be crucial in ensuring this.

The left-hand side of Eq.~\eqref{eq:IBP_test} is very easy to compute in the case of circular orbits. First, we compute the term inside the explicit time derivative. It only requires the expression of the multipolar moments (bearing time derivatives) at Newtonian order. Namely, we require 
\begin{subequations}\begin{align}
    \dM_{ij}^{(3)} &= -\frac{8 \nu c^{5} y^{5/2}}{G} n^{\langle i} \lambda^{j \rangle} +\mathcal{O}\left(\frac{1}{c^2}\right) \,,\\*
    \dM_{ij}^{(4)} &= \frac{8 \nu c^8  y^4}{G^2 m}\left(n^{\langle i} n^{j \rangle}- \lambda^{\langle i} \lambda^{j \rangle}\right)+ \mathcal{O}\left(\frac{1}{c^2}\right)   \,, 
\end{align}\end{subequations}
where the symmetric trace-free projection is denoted with angled brackets around the indices, namely \mbox{$a^{\langle i} b^{j\rangle} \equiv \frac{1}{2}\left(a^i b^j +   a^j b^i\right)-\frac{1}{3}\delta^{ij} a^k b^k$}.
For the purpose of evaluating the tail integral, we can then project the corotating basis $(\bm{n},\bm{\lambda})$ onto a time-independent basis $(\bm{n}_0,\bm{\lambda}_0)$ using
\begin{subequations}
\begin{align}
    \bm{n} = \cos \phi \,\bm{n}_0 + \sin \phi\, \bm{\lambda}_0 \,, \\*
    \bm{\lambda} = -\sin \phi \,\bm{n}_0 + \cos \phi \, \bm{\lambda}_0 \,,
\end{align}\end{subequations}
where at leading order the phase reads \mbox{$\phi(t \pm \tau) = \phi(t) \pm \omega\tau + ...$} 
Using only these Newtonian expressions alongside the well-known adiabatic integration formula~\eqref{eq:adiabatic_integration},  I finally find that 
\begin{align}\label{eq:term_inside_time_derivative_IBP}
    &\dM^{(3)}_{ij} \int_0^{\infty}\dd\tau \ln\left(\frac{c \tau}{2 b_0}\right) \dM^{(4)}_{ij}(t\pm\tau)\nonumber\\*
    &= \pm \frac{16 c^{10} \nu^2 y^5}{G^2}\bigg[2 \gamma_{\text{E}} + \ln(16) - 3 \ln(y) - 2 \ln\left(\frac{G m}{c^2 b_0}\right)\bigg] .
\end{align}
The latter expression is independent of the orbital phase, and is approximately constant. Its only time dependence is in the slowly evolving frequency, whose time evolution is given at leading 2.5PN order by
\begin{align}\label{eq:dxdt_leading}
    \frac{\dd y}{\dd t} = \frac{64 c^3 \nu y^5}{5 G m}\,,
\end{align}
where we recall that $y$ is defined by Eq.~\eqref{eq:y_def}.
Thus, I~find that at leading order, 
\begin{align}\label{eq:IBP_test_LHS}
    \frac{\dd}{\dd t}&\Bigg[\dM^{(3)}_{ij} \int_0^{\infty}\dd\tau \ln\left(\frac{c \tau}{2 b_0}\right) \dM^{(4)}_{ij}(t\pm\tau)\Bigg] \nonumber\\*
    &= \pm \frac{1024 c^{13} \nu^3 y^9}{5 G^3 m}\Bigg[3+10\gamma_\text{E} + 5 \ln(16) \nonumber\\*
    &\qquad\qquad\qquad\quad  + 15 \ln(y) - 10\ln\left(\frac{G m}{c^2 b_0}\right)\Bigg]\,.
\end{align}

The right-hand side of Eq.~\eqref{eq:IBP_test} is much more subtle to compute. Indeed, I found that it exactly vanishes if we naively plug in the Newtonian values for the moments and use the adiabatic integration formula, because of the tensorial structure of the multipolar moments. This is linked to the fact that \eqref{eq:term_inside_time_derivative_IBP} is approximately constant. In order to find that leading nonzero term in the right-hand side of \eqref{eq:IBP_test}, we need to use the 2.5PN-accurate expression of the multipolar moments, which are given in the Appendix~\ref{app:quadrupole}. These include crucial 2.5PN corrections to the moments, which have a different tensorial structure because they arise from radiation reaction effects. Most importantly, in order to find the correct result, we also need to use the postadiabatic integration formulas
\eqref{eq:postadiabatic_integration}. Only then do we recover that 
\begin{align}\label{eq:IBP_test_RHS}
    &\Big(\text{RHS of \eqref{eq:IBP_test}}\Big)\nonumber= \pm \frac{1024 c^{13} \nu^3 y^9}{5 G^3 m}\Bigg[3+10\gamma_\text{E} + 5 \ln(16) \nonumber\\*
    &\qquad\qquad\qquad\qquad\qquad\qquad\! + 15 \ln(y) - 10\ln\left(\frac{G m}{c^2 b_0}\right)\!\Bigg]\,,
\end{align}
which concludes the verification.

Let me insist that if we were to use a postadiabatic integration formula different from the one derived in Eq.~\eqref{eq:postadiabatic_integration} (e.g., changing the signs of the postadiabatic contributions), we would not recover the correct answer for Eq.~~\eqref{eq:IBP_test_RHS}, namely it would not match Eq.~\eqref{eq:IBP_test_LHS}. Since the left-hand side requires none of this information, this is a strong check of our integration formulas. I~have repeated this test successfully with identities similar to Eq.~\eqref{eq:IBP_test}, but with other combinations of time derivatives.

\section{Computation of the Schott terms for circular orbits}
\label{sec:schott_circular}
All the tools are now available to compute the pseudo-Schott term \eqref{seq:E_diss_4PN} for circular orbits. Since we are working at leading post-Newtonian order, we can safely replace the ADM mass by the total mass, $\dM = m + \mathcal{O}(x)$. Using simply the Newtonian moments, we trivially find that 
\begin{subequations}\begin{align}
    \dM_{ij}^{(3)} \dM_{ij}^{(3)} &= \frac{32c^{10} \nu^2 y^5}{G^2} \,,\\*
    \epsilon_{ijk}\dM_{jl}^{(2)} \dM_{kl}^{(3)} &= \frac{16 c^7 m \nu^2 y^{7/2}}{G} \ell_0^i \,,
\end{align}\end{subequations}
where $\ell_0^i = \epsilon_{ijk}\, n_0^i \lambda_0^j$ completes the right-handed orthonormal tetrad $(\bm{n}_0,\bm{\lambda}_0,\bm{\ell}_0)$.
  
I have found explicitly that terms which are of the type ``retarded plus advanced'' all vanish identically at leading order:
\begin{align}
     \dM_{ij}^{(1)} \!\!\int_{0}^\infty \!\!\!  \dd\tau \ln\left(\frac{c\tau}{2 b_0}\right) \!\left[\dM_{ij}^{(6)}(t-\tau) +
    \dM_{ij}^{(6)}(t+\tau) \right] &= 0 \,, \nonumber\\
    \dM_{ij}^{(2)} \!\!\int_{0}^\infty \!\!\!  \dd\tau \ln\left(\frac{c\tau}{2 b_0}\right) \!\left[\dM_{ij}^{(5)}(t-\tau) +
    \dM_{ij}^{(5)}(t+\tau) \right] &= 0 \,,\nonumber\\
    \epsilon_{ijk}\dM_{jl}\!\! \int_{0}^\infty \!\!\! \dd\tau \ln\left(\frac{c\tau}{2 b_0}\right)\! \left[\dM_{kl}^{(6)}(t-\tau) +
    \dM_{kl}^{(6)}(t+\tau) \right] &= 0 \,,\nonumber\\
    \epsilon_{ijk}\dM_{jl}^{(1)}\!\! \int_{0}^\infty \!\!\! \dd\tau \ln\left(\frac{c\tau}{2 b_0}\right)\! \left[\dM_{kl}^{(5)}(t-\tau) +
    \dM_{kl}^{(5)}(t+\tau) \right] &= 0 \,.
\end{align}
We only need these identities at leading order, but recall that these integrals are indeed nonvanishing starting at relative 2.5PN order. 
Using again the Newtonian moments and the adiabatic integration formulas, we then find that the integrals of the type ``retarded minus advanced'' or simply advanced read
\begin{subequations}\begin{align}
    &\dM_{ij}^{(3)} \int_{0}^\infty  \dd\tau \ln\left(\frac{c\tau}{2 b_0}\right) \left[\dM_{ij}^{(4)}(t-\tau) -  \dM_{ij}^{(4)}(t+\tau) \right] \nonumber\\*
    &= 	-\frac{32 c^{10} \nu^2 y^5 }{G^2}\Bigg[3 \ln y +2  \gamma_\text{E} +4 \ln 2-2 \ln \left(\frac{G m}{c^2 b_0}\right)\! \Bigg] \,, \nonumber\\*
    & \\
    & \epsilon_{ijk}  \dM_{jl}^{(2)} \int_{0}^\infty  \dd\tau \ln\left(\frac{c\tau}{2 b_0}\right) \dM_{kl}^{(4)}(t+\tau)  \nonumber\\*
    &=\frac{8 c^7 m \nu^2 y^{7/2}}{G}\Bigg[3 \ln y +2 \gamma_\text{E} +4 \ln 2-2 \ln \left(\frac{G m}{c^2 b_0}\right)\! \Bigg] \ell_0^i \,.\nonumber\\*
    & 
\end{align}\end{subequations}

Finally, we treat the terms akin to  genuine tails  of memory~\cite{TB23_ToM}, i.e., the first line of Eqs.~\eqref{seq:E_diss_4PN} and \eqref{seq:J_diss_4PN}. We first notice that the integrand of the $\rho$-integral vanishes if we use the Newtonian moments and the adiabatic integration formula \eqref{eq:adiabatic_integration}, as discussed in Sec.~\ref{subsec:postadiabatic_derivation}. 
To obtain the leading order nonvanishing contribution, we will in fact require the 2.5PN-accurate moments (given in Appendix~\ref{app:quadrupole}) and the postadiabatic integration formula~\eqref{eq:postadiabatic_integration}. We then find that 

\begin{widetext}
 
\begin{subequations}\begin{align}\label{seq:E_diss_4PN}
    &\dM_{ij}^{(4)}(t) \int_{0}^\infty \dd \tau \ln\left(\frac{c\tau}{2 b_0}\right) \left[\dM_{ij}^{(4)}(t-\tau) - \dM_{ij}^{(4)}(t+\tau)\right]
    = -\frac{1024 c^{13} \nu^3 y^9}{5
   G^3 m} \Bigg[15 \ln y+10 \gamma_\text{E} -7+20 \ln 2
   -10 \ln \left(\frac{G m}{c^2 b_0}\right)\Bigg]\,, \\
   & \dM_{jl}^{(3)}\int_{0}^\infty \dd \tau \ln\left(\frac{c\tau}{2 b_0}\right) \left[\dM_{kl}^{(4)}(t-\tau) - \dM_{kl}^{(4)}(t+\tau)\right]
   = \frac{3584 c^{10} \nu^3 y^{15/2}}{5 G^2} \ell_0^i \,.
\end{align}\end{subequations}
   
\end{widetext}

Then, we need to perform an integration with respect to time on this expression, but we notice that it only depends on time through $x$, which evolves on radiation-reaction timescales. Thus, we perform the time integration as follows:
\begin{align}
    \int_0^\infty \dd \rho f\big(y(t-\rho)\big) &= \int_{0}^y \dd y' \frac{f(y')}{\dd y' / \dd t} \,,
\end{align}
where $\dd y' / \dd t$ is given at the required order in Eq.~\eqref{eq:dxdt_leading}. Using this, I find
\begin{widetext}
\begin{subequations}\begin{align}
    &\int_0^\infty   \dd \rho \, \dM_{ij}^{(4)}(t-\rho) \int_{0}^\infty    \dd \tau \ln\left(\frac{c\tau}{2 b_0}\right)\! \Big[\dM_{ij}^{(4)}(t-\rho-\tau)  - \dM_{ij}^{(4)}(t-\rho+\tau)\Big]\nonumber\\*
    &\qquad  = -\frac{16 c^{10} \nu^2 y^5 }{G^2}\Bigg[3 \ln y +2 \gamma_\text{E} -2+ 4\ln 2
    -2 \ln \left(\frac{G m}{c^2 b_0}\right) \Bigg]\,, \\
    &\epsilon_{ijk} 
     \int_0^\infty  \dd \rho \, \dM_{jl}^{(3)}(t-\rho) \int_{0}^\infty  \dd \tau \ln\left(\frac{c\tau}{2 b_0}\right) \left[\dM_{kl}^{(4)}(t-\rho-\tau) - \dM_{kl}^{(4)}(t-\rho+\tau)\right]\nonumber\\*
    &\qquad = \frac{16 c^7 m \nu^2 y^{7/2}}{G} \ell_0^i \,.
\end{align}\end{subequations}
Adding it all together, I obtain finally
\begin{subequations}\label{eq:EJ_diss_4PN_circ}\begin{align}\label{seq:E_diss_4PN_circ}
      E^{\text{circ}}_{\text{diss}}  &=  - \frac{c^2 m \nu^2 y^5}{2} \Bigg[\frac{128}{5}   \ln\left(\frac{G m}{c^2 b_0}\right) -\frac{192}{5} \ln(y)   -\frac{128}{5} \gamma_{\mathrm{E}}  -\frac{256}{5}  \ln(2) -\frac{32}{15}\Bigg] \,,\\
      J^{\text{circ}}_{\text{diss}}  &=  \frac{G m^2 \nu^2 y^{7/2}}{c} \Bigg[-\frac{64}{5}   \ln\left(\frac{G m}{c^2 b_0}\right) +\frac{96}{5} \ln(y)   + \frac{64}{5} \gamma_{\mathrm{E}} +\frac{128}{5}  \ln(2)  +\frac{16}{15}\Bigg] \,.
\end{align}\end{subequations}
\end{widetext}
One immediately notices that these corrections satisfy at this order
\begin{align}
     E^{\text{circ}}_{\text{diss}} &= \omega \, J^{\text{circ}}_{\text{diss}} \,.
\end{align}
We shall see later that this is a necessary condition for the consistency of our calculation.

\section{Complete energy and angular momentum}
\label{sec:circular_binding_energy}

Now, recall that the conservative energy and angular momentum are given at 4.5PN as follows \cite{BBFM17}:
\begin{widetext}
\begin{subequations}
    \begin{align}\label{eq:cons_energy_y}
	E^\text{circ}_\text{cons}(y) &= -\frac{m \nu c^2 y}{2} \Biggl\{ 1 + y\biggl( - \frac{3}{4} -
	\frac{\nu}{12} \biggr)  + y^2\biggl( - \frac{27}{8} +
	\frac{19}{8} \nu - \frac{\nu^2}{24} \biggr)  \nonumber\\
	&\qquad\quad + y^3\biggl[ - \frac{675}{64} + \nu\biggl(
	\frac{34445}{576} - \frac{205}{96} \pi^2 \biggr)  -
	\frac{155}{96} \nu^2 - \frac{35}{5184} \nu^3 \biggr] 
	\nonumber\\ 
&\qquad\quad + y^4\Biggl[ - \frac{3969}{128} +
	\nu\biggl(-\frac{123671}{5760}+\frac{9037}{1536}\pi^2 +
	\frac{896}{15}\gamma_\text{E}+ \frac{448}{15} \ln(16
	y)\biggr)\nonumber\\
& \qquad\quad\qquad +
	\nu^2\biggl(-\frac{498449}{3456}+\frac{3157}{576}\pi^2\biggr)
	+\frac{301}{1728}\nu^3 + \frac{77}{31104}\nu^4\Biggr]  +
	\mathcal{O}\bigl(y^5\bigr)
	\Biggr\} \,,\\
    J^\text{circ}_\text{cons}(y) &= \frac{G m^2 \nu}{c \sqrt{y}}\Biggr\{1 + y \left(\frac{3}{2}+\frac{\nu }{6}\right) + y^2\left(\frac{27}{8} -\frac{19 \nu }{8}+\frac{\nu^2}{24}\right)
    \nonumber\\
   &\qquad\quad +  y^3\Biggl[\frac{135}{16} +\left(-\frac{6889}{144}+\frac{41}{24}\pi^2\right) \nu  + \frac{31}{24}\nu^2 + \frac{7}{1296}\nu^3\Biggl] \nonumber\\
   &\qquad\quad +  y^4 \Biggl[\frac{2835}{128} +\nu  \left(\frac{98869}{5760} -\frac{6455 \pi^2}{1536}-\frac{128 \gamma_{\mathrm{E}}
   }{3} -\frac{64}{3} \log (16 y) \right) \nonumber\\
   & \qquad\quad\qquad+\nu^2\left(\frac{356035}{3456}-\frac{2255 \pi
^2}{576}\right)  -\frac{215}{1728}\nu^3-\frac{55}{31104} \nu^4 \Biggl] +
	\mathcal{O}\bigl(y^5\bigr)
	\Biggr\}\,.
\end{align}\end{subequations}
\end{widetext}
It is easy to verify that these satisfy the thermodynamic relation (masses kept fixed) \cite{LBW12}
\begin{align}
    \frac{\dd E^\text{circ}_\text{cons}}{\dd\omega} = \omega\, \frac{\dd J^\text{circ}_\text{cons}}{\dd \omega}\,,
\end{align}
where $\omega$ is the orbital frequency entering \eqref{eq:y_def}. One can then obtain the binding energy and angular momentum as 
\begin{subequations}\label{eq:EJ_binding_circ_y}\begin{align}\label{subeq:E_binding_circ_y}
    E^\text{circ}(y;b_0) &= E^\text{circ}_\text{cons}(y) + E^\text{circ}_\text{diss}(y;b_0) \,, \\* 
    \label{subeq:J_binding_circ_y}
    J^\text{circ}(y;b_0) &= J^\text{circ}_\text{cons}(y) + J^\text{circ}_\text{diss}(y;b_0) \,,
\end{align}\end{subequations}
which depends manifestly on our choice of slicing through the constant $b_0$.
Now, the orbital frequency is a very important parameter, but it is not an observable quantity which can be measured by an asymptotic observer at future null infinity. Instead, the main observable is the half-frequency of the $(\ell,\mathrm{m})=(2,2)$ mode appearing in the asymptotic gravitational wave, which we denote $\Omega_{22}$. We then introduce  the associated dimensionless PN counting parameter,
\begin{align}
    x = \left(\frac{G m \Omega_{22}}{c^3}\right)^{2/3}\,,
\end{align}
which differs from the orbital frequency $y$ by a 4PN correction, namely~\cite{BFHLTa,BFHLTb}:
\begin{align}\label{eq:x_of_y}
x &=y  \Bigg\{1+y^4 \nu \Bigg[\frac{128}{5}   \ln\left(\frac{G m}{c^2 b_0}\right) -\frac{192}{5} \ln(y) \nonumber\\*
&\qquad\qquad\qquad -\frac{128}{5} \gamma_{\mathrm{E}}  -\frac{32}{15}-\frac{256}{5}  \ln(2)\Bigg]\Bigg\} \,.
\end{align}
Notice the structure of the 4PN correction: the coefficients are exactly the same as the Schott term \eqref{seq:E_diss_4PN_circ}. This observation plays a crucial role. Indeed, if we  now take $E(y;b_0)$ as given by \eqref{subeq:E_binding_circ_y} and replace every $y$ by its expression in terms of $x$, we find the following expression for the energy in terms of $x$:

\begin{widetext}
    \begin{align}\label{eq:total_energy_x}
	E^\text{circ}(x) &= -\frac{m \nu c^2 x}{2} \Biggl\{ 1 + x\biggl( - \frac{3}{4} -
	\frac{\nu}{12} \biggr)  + x^2\biggl( - \frac{27}{8} +
	\frac{19}{8} \nu - \frac{\nu^2}{24} \biggr)  \nonumber\\
	&\qquad\quad + x^3\biggl[ - \frac{675}{64} + \nu\biggl(
	\frac{34445}{576} - \frac{205}{96} \pi^2 \biggr)  -
	\frac{155}{96} \nu^2 - \frac{35}{5184} \nu^3 \biggr] 
	\nonumber\\ 
&\qquad\quad + x^4\Biggl[ - \frac{3969}{128} +
	\nu\biggl(-\frac{123671}{5760}+\frac{9037}{1536}\pi^2 +
	\frac{896}{15}\gamma_\text{E}+ \frac{448}{15} \ln(16
	x)\biggr)\nonumber\\
& \qquad\quad\qquad +
	\nu^2\biggl(-\frac{498449}{3456}+\frac{3157}{576}\pi^2\biggr)
	+\frac{301}{1728}\nu^3 + \frac{77}{31104}\nu^4\Biggr]  +
	\mathcal{O}\bigl(x^5\bigr)
	\Biggr\} \,, \\
    J^\text{circ}(x) &= \frac{G m^2 \nu}{c \sqrt{x}}\Biggr\{1 + x \left(\frac{3}{2}+\frac{\nu }{6}\right) + x^2\left(\frac{27}{8} -\frac{19 \nu }{8}+\frac{\nu^2}{24}\right)
    \nonumber\\
   &\qquad\quad +  x^3\Biggl[\frac{135}{16} +\left(-\frac{6889}{144}+\frac{41}{24}\pi^2\right) \nu  + \frac{31}{24}\nu^2 + \frac{7}{1296}\nu^3\Biggl] \nonumber\\
   &\qquad\quad +  x^4 \Biggl[\frac{2835}{128} +\nu  \left(\frac{98869}{5760} -\frac{6455 \pi^2}{1536}-\frac{128 \gamma_{\mathrm{E}}
   }{3} -\frac{64}{3} \log (16 x) \right) \nonumber\\
   & \qquad\quad\qquad+\nu^2\left(\frac{356035}{3456}-\frac{2255 \pi^2}{576}\right)  -\frac{215}{1728}\nu^3-\frac{55}{31104} \nu^4 \Biggl] +
	\mathcal{O}\bigl(x^5\bigr)
	\Biggr\}\,.
\end{align}
\end{widetext}
The first observation is that the dependence in $b_0$ has dropped out. The second observation is that, at least at 4.5PN order,  the functional expression of $E$ in terms of~$x$ [given by Eq.~\eqref{eq:total_energy_x}] is exactly the same as $E_\mathrm{cons}$ in terms of~$y$ [given by Eq.~\eqref{eq:cons_energy_y}]. This observation was the key heuristic postulate used in \cite{BFHLTa, BFHLTb} to derive the phasing, see the paragraph following Eq.~(6.13) in Ref.~\cite{BFHLTb}. An immediate consequence if that the binding energy and angular momentum, which include both conservative and dissipative contributions, also satisfy a thermodynamic relation (masses kept fixed), but now with respect to the \textit{waveform} frequency
\begin{align}\label{eq:thermo_relation_EJ}
    \frac{\dd E^\text{circ}}{\dd\Omega_{22}} = \Omega_{22}\, \frac{\dd J^\text{circ}}{\dd \Omega_{22}} \,.
\end{align}
This expression is reminiscent of a dissipative first law of black hole binary mechanics~\cite{GLTP25}, and is consistent with Ref.~\cite{Khairnar:2024rzs}, as we will see later on.
\newpage

\section{Impact on frequency chirp and phasing}
\label{sec:phasing}

\subsection{In terms of the orbital frequency}
\label{subsec:phasing_orbital}

The flux of energy for circular orbits was given  at 4.5PN in terms of the waveform frequency in Refs.~\cite{BFHLTa, BFHLTb}, and in terms of the orbital frequency in (VI.61) of Ref.~\cite{Trestini:2023khz}.
The flux of angular momentum of was subsequently computed in Ref.~\cite{Khairnar:2024rzs} by replacing the radiative moments of Eq.~\eqref{seq:flux_J_UL_VL} by their expressions in terms of the canonical moments given in Ref.~~\cite{BFHLTb}, whose explicit expressions for circular orbits are also given in~\cite{BFHLTb}.
There is confusion in Ref.~\cite{Khairnar:2024rzs} whether $\Omega$ denotes orbital or waveform frequency, but it was confirmed to me in private correspondence with one of the authors (Aniket Khairnar) that  Eq.~(20) of Ref.~\cite{Khairnar:2024rzs}
is given in terms of the \textit{waveform} frequency $x = (Gm\Omega_{22}/c^3)^{2/3}$.  For the present work, I simply reexpressed their result in terms of orbital frequency using Eq.~\eqref{eq:x_of_y}. A strong test of their result is that they found that their angular momentum flux was proportional to our energy flux through 4.5PN order, with the proportionality factor being the \textit{waveform} frequency, see Eq.~\eqref{eq:relation_fluxes_waveform_frequency} below. This is consistent with the thermodynamic observation we obtained in Eq.~\eqref{eq:thermo_relation_EJ}.

Thus, we recall the expressions for the  fluxes of energy and angular momentum at future null infinity in case of circular orbits~\cite{BFHLTa,BFHLTb,Khairnar:2024rzs}
\begin{widetext}
\begin{subequations}\label{eq:flux_EJ_y}\begin{align}\label{seq:flux_E_y}
\mathcal{F}^\infty_E(y;b_0) &= \frac{32c^5}{5G}\nu^2 y^5 \Biggl\{
1 +  y \biggl(-\frac{1247}{336} - \frac{35}{12}\nu \biggr) 
+ 4\pi y^{3/2}
+ y^2\biggl(-\frac{44711}{9072} +\frac{9271}{504}\nu + \frac{65}{18} \nu^2\biggr)  + \pi y^{5/2}\biggl(-\frac{8191}{672}-\frac{583}{24}\nu\biggr)
\nonumber \\
& 
+ y^3 \Biggl[\frac{6643739519}{69854400}+ \frac{16}{3}\pi^2-\frac{1712}{105}\gamma_\text{E} - \frac{856}{105} \ln (16 y) 
 + \nu\biggl(-\frac{134543}{7776} + \frac{41}{48}\pi^2 \biggr) 
- \frac{94403}{3024}\nu^2 
- \frac{775}{324}\nu^3 \Biggr] 
\nonumber\\
&
+ \pi y^{7/2} \biggl(-\frac{16285}{504} + \frac{214745}{1728}\nu +\frac{193385}{3024}\nu^2\biggr) 
\nonumber\\
&
+ y^4\Biggl[ -\frac{323105549467}{3178375200} + \frac{232597}{4410}\gamma_\text{E} - \frac{1369}{126} \pi^2 
+ \frac{39931}{294}\ln 2 - \frac{47385}{1568}\ln 3 + \frac{232597}{8820}\ln y   
\nonumber\\
& \quad
+ \nu\Biggl( -\frac{1467849789229}{1466942400} + \frac{10118}{245}\gamma_\text{E} - \frac{267127}{4608}\pi^2
- \frac{85418}{2205}\ln 2 + \frac{47385}{392}\ln 3  - \frac{26301}{245}\ln y + 128 \ln\left(\frac{G m}{c^2 b_0 }\right) \Biggr)
\nonumber\\*
& \qquad
+ \nu^2\biggl( \frac{1607125}{6804} - \frac{3157}{384}\pi^2 \biggr) + \frac{6875}{504}\nu^3 + \frac{5}{6}\nu^4 \Biggr] 
\nonumber\\
& 
+ \pi y^{9/2} \Biggl[ \frac{265978667519}{745113600} - \frac{6848}{105}\gamma_\text{E} - \frac{3424}{105} \ln (16  y)
+ \nu\biggl( \frac{2062241}{22176} + \frac{41}{12}\pi^2 \biggr)
\nonumber\\ 
& \qquad\quad
- \frac{133112905}{290304}\nu^2 - \frac{3719141}{38016}\nu^3 \Biggr] 
+ \mathcal{O}\bigl(x^5\bigr) \Biggr\}\,,\\
\label{seq:flux_J_y}
\mathcal{F}^\infty_J(y;b_0) &= \frac{32c^2 m}{5}\nu^2 y^{7/2} \Biggl\{
1 +  y \biggl(-\frac{1247}{336} - \frac{35}{12}\nu \biggr) 
+ 4\pi y^{3/2}
+ y^2\biggl(-\frac{44711}{9072} +\frac{9271}{504}\nu + \frac{65}{18} \nu^2\biggr)  + \pi y^{5/2}\biggl(-\frac{8191}{672}-\frac{583}{24}\nu\biggr)
\nonumber \\*
& 
+ y^3 \Biggl[\frac{6643739519}{69854400}+ \frac{16}{3}\pi^2-\frac{1712}{105}\gamma_\text{E} - \frac{856}{105} \ln (16 y) 
 + \nu\biggl(-\frac{134543}{7776} + \frac{41}{48}\pi^2 \biggr) 
- \frac{94403}{3024}\nu^2 
- \frac{775}{324}\nu^3 \Biggr] 
\nonumber\\
&
+ \pi y^{7/2} \biggl(-\frac{16285}{504} + \frac{214745}{1728}\nu +\frac{193385}{3024}\nu^2\biggr) 
\nonumber\\
&
+ y^4\Biggl[ -\frac{323105549467}{3178375200} + \frac{232597}{4410}\gamma_\text{E} - \frac{1369}{126} \pi^2 
+ \frac{39931}{294}\ln 2 - \frac{47385}{1568}\ln 3 + \frac{232597}{8820}\ln y   
\nonumber\\*
& \quad
+ \nu\Biggl( -\frac{1463155573549}{1466942400} + \frac{19526}{245}\gamma_\text{E} - \frac{267127}{4608}\pi^2
+ \frac{83926}{2205}\ln 2 + \frac{47385}{392}\ln 3  - \frac{12189}{245}\ln y + \frac{448}{5} \ln\left(\frac{G m}{c^2 b_0 }\right) \Biggr)
\nonumber\\*
& \qquad
+ \nu^2\biggl( \frac{1607125}{6804} - \frac{3157}{384}\pi^2 \biggr) + \frac{6875}{504}\nu^3 + \frac{5}{6}\nu^4 \Biggr] 
\nonumber\\
& 
+  \pi y^{9/2} \Biggl[ \frac{265978667519}{745113600} - \frac{6848}{105}\gamma_\text{E} - \frac{3424}{105} \ln (16  y)
+ \nu\biggl( \frac{2062241}{22176} + \frac{41}{12}\pi^2 \biggr)
\nonumber\\* 
& \qquad\quad
- \frac{133112905}{290304}\nu^2 - \frac{3719141}{38016}\nu^3 \Biggr]
+ \mathcal{O}\bigl(x^5\bigr) \Biggr\}\,.
\end{align}\end{subequations}
%
Note that these two fluxes do not only differ by a prefactor: differences appear at 4PN, namely
\begin{align}
    &\mathcal{F}^\infty_{E}(y)- \omega\,  \mathcal{F}^\infty_{J}(y)  = \frac{32 c^5 y^9 \nu^3}{5G}\bigg[ \frac{192}{5} \ln\left(\frac{G m}{c^2 b_0}\right) - \frac{288}{5}\ln(y)  - \frac{16}{5} - \frac{192}{5}\gamma_E  - \frac{384}{5}\ln(2) \bigg]
\end{align}
\end{widetext}
As we will see, this difference arises because the ratio between the two fluxes should instead be the \textit{waveform} frequency~\cite{Khairnar:2024rzs}, see~Eq.~\eqref{eq:relation_fluxes_waveform_frequency}. 

In the case of binary black holes, one also should account for the fact that horizon flux contributions enter at 4PN. In the case of other compact objects such as neutron stars, the analogous contribution is tidal heating, which also enters formally at 4PN order when one assumes that the compactness \cite{Ozel:2016oaf} of these objects is of order $1$. We find that the horizon fluxes read 
\begin{subequations}\label{eq:EJ_flux_horizon_y}
    \begin{align}\label{subeq:E_flux_horizon_y}
    F_{E}^{\mathcal{H}} &= \frac{32 c^5 \nu^2}{5 G} \aleph_8\, y^9 \,,\\ \label{subeq:J_flux_horizon_y}
    F_{J}^{\mathcal{H}} &= \frac{32 c^2 m \nu^2}{5} \aleph_8 \,y^{15/2}\,,
\end{align}
\end{subequations}
where I have define $\aleph_{2n}$ as the  tidal dissipation (or black hole absorption) coefficient entering at the $n$PN order beyond the leading, quadrupolar flux (``aleph'' stands for ``absorption''). Note that for nonspinning binary black holes, one has $\aleph_8 = 1-4\nu +2\nu^2$ \cite{Taylor:2008xy, Nagar:2011aa}; see also \cite{Saketh:2024juq} for tidal heating computations for neutron stars.

We now have all the ingredients entering the flux balance laws, which read
\begin{subequations}\label{eq:flux_balance_laws}
\begin{align}
    \frac{\dd E}{\dd t} &= - \mathcal{F}_E \,,\\*
    \frac{\dd J}{\dd t} &= - \mathcal{F}_J \,,
\end{align}\end{subequations}
where we have reintroduced the horizon fluxes: \mbox{$\mathcal{F}_E = \mathcal{F}^\infty_E+\mathcal{F}_E^{\mathcal{H}}$} and \mbox{$\mathcal{F}_J = \mathcal{F}^\infty_J+\mathcal{F}_J^{\mathcal{H}}$}.
Using the chain rule, we find that the differential equation governing the orbital frequency evolution reads    
\begin{align}
    \frac{\dd y}{\dd t} &= - \frac{\mathcal{F}_E(y;b_0)}{\dd E(y;b_0) / \dd y} \,,
\end{align}
where we are ignoring the mass evolution driven by black hole absorption (these lead to 5PN effects in the phasing which are negligible here \cite{Warburton:2024xnr}).
Crucially, we find that $b_0$ drops out from the final results at 4.5PN, as one would expect physically. This observation is nontrivial: the $b_0$ dependence in the energy was obtained only with the knowledge of the 4PN tail acceleration and the  1.5PN flux, whereas the  $b_0$ appears at 4PN in the flux, and arises from various parts of the computation. The explicit expression of the right-hand side of this equation is most conveniently split into a contribution from the fluxes at infinity and another subdominant contribution from the horizon fluxes
\begin{align}\label{eq:dydt_total}
    \frac{\dd y}{\dd t} &= \left\{\frac{\dd y}{\dd t}\right\}_{\infty} + \left\{\frac{\dd y}{\dd t}\right\}_{\mathcal{H}} \,.
\end{align}
The main contribution due to the flux at infinity reads
\begin{align}\label{eq:dydt_infinity}
    \left\{\frac{\dd y}{\dd t}\right\}_{\infty} &= \frac{c^3}{G m}\frak{D}(y;\nu)\,,
\end{align}
where the right-hand side is driven by the function 
\begin{widetext}
\begin{align} \label{eq:D_of_z}
\frak{D}(z;\nu) &= \frac{64  \nu  z^5}{5} \Bigg\{1 +         z\left(-\frac{743}{336}-\frac{11}{4}\nu\right)
   +4 \pi    z^{3/2}
   +z^2\left(\frac{34103}{18144}+\frac{13661}{2016}\nu +\frac{59}{18} \nu^2 \right)
   +\pi z^{5/2} \left(-\frac{4159 }{672} -\frac{189}{8}\nu \right) \nonumber\\*
   &\qquad +z^3 \Bigg[\frac{16447322263}{139708800}+\frac{16}{3}\pi^2-\frac{1712}{105}\gamma_\text{E}   -\frac{856}{105} \log (16 z) +\left(-\frac{56198689}{217728}+\frac{451}{48} \pi^2\right) \nu
   +\frac{541}{896} \nu^2
   -\frac{5605}{2592}\nu^3
   \Bigg] \nonumber\\
   &\qquad + \pi z^{7/2}\Bigg[ -\frac{4415}{4032} +\frac{358675}{6048} \nu + \frac{91495}{1512}\nu^2\Bigg]  \nonumber\\
   &\qquad +z^4 \Bigg[\frac{3959271176713}{25427001600}  -\frac{361}{126}\pi^2+\frac{124741}{4410}\gamma_\text{E}-\frac{47385}{1568}\ln
   (3)+\frac{127751}{1470} \ln (2)
    +\frac{124741}{8820}\ln(z)
    \nonumber\\*
   & \qquad\qquad 
   +\nu  \left(-\frac{302567610617}{391184640}-\frac{1472377}{16128}\pi^2 -\frac{58250}{441}\gamma_\text{E} +\frac{47385 \log (3)}{392}-\frac{850042 \log (2)}{2205}-\frac{29125 \log (z)}{441}\right) \nonumber\\*
   &\qquad\qquad 
   +\nu^2\left(\frac{504221849}{326592}-\frac{22099}{384} \pi
^2\right) 
   -\frac{1909807}{62208} \nu^3
   +\frac{917}{1152} \nu^4
   \Bigg]  \nonumber\\
   & \qquad+ \pi z^{9/2} \Bigg[ \frac{343801320119}{745113600} 
   -\frac{6848}{105} \gamma_\text{E} -\frac{3424}{105} \log (16 z)
\nonumber\\*
& \qquad\qquad+\nu\left(-\frac{516333533}{532224}+\frac{451}{12}\pi^2\right) 
-\frac{6821669}{41472}\nu^2
    -\frac{24107249}{266112}\nu^3
   \Bigg]  \Bigg\}
\end{align}
\end{widetext}
The contributions due to the horizon fluxes are simply given by
\begin{align} \label{eq:dydt_horizon}
\left\{\frac{\dd y}{\dd t}\right\}_{\mathcal{H}} &= \frac{64 c^3 \nu }{5 G m}   \aleph_8\, y^{9}\,.
\end{align}
I have checked explicitly that, as expected, using the flux balance law for the angular momentum yields the same result.

The result of Eqs.~\eqref{eq:dydt_total}~contains all the needed physical information, but it is nonetheless useful to integrate it analytically, in a post-Newtonian sense, to get (i) the orbital phase $\phi$ in terms of the orbital frequency $y$, and (ii) the orbital frequency $y$ as an explicit function of time.
First, one can express time in terms of $y$ through the relation
\begin{align} \label{eq:t_inTermsOf_y}
    t - t_c = \int^{y} \frac{\dd y'}{dy'/dt}\,,
\end{align}
where $d y '/dt$ is given in terms of $y'$ in Eq.~\eqref{eq:dydt_total} and where $t_c$ is an arbitrary integration constant which can be chosen to simplify the expression, and will be interpreted loosely as the ``coalescence time''. Performing a post-Newtonian expansion on the integrand, one can straightforwardly perform the integral of Eq.~\eqref{eq:t_inTermsOf_y}. It is then straightforward to invert the relation in a post-Newtonian sense, so as to express the orbital frequency in terms of time. For this, it is convenient to introduce the dimensionless time variable
\begin{align}
    \tau = \frac{\nu c^3}{5 G m}(t_c-t) \,.
\end{align}
We thus find that the orbital frequency evolves as 
\begin{widetext}
\begin{align} 
	y(\tau)   &= \frac{\tau^{-1/4}}{4}\Biggl\{ 1 + \tau^{-1/4} \biggl( \frac{743}{4032} +
	\frac{11}{48}\nu\biggr) - \frac{\pi}{5}\,\tau^{-3/8}
	\nonumber \\ & \quad \quad 
    + \tau^{-1/2}\biggl( \frac{19583}{254016} +
	\frac{24401}{193536} \nu + \frac{31}{288} \nu^2 \biggr)
	 +  \pi \,\tau^{-5/8} \biggl(-\frac{11891}{53760} +
	\frac{109}{1920}\nu\biggr) 
	\nonumber \\ & \quad
	\quad + \tau^{-3/4} \Biggl[-\frac{10052469856691}{6008596070400} +
	\frac{\pi^2}{6} + \frac{107}{420}\gamma_\text{E} -
	\frac{107}{3360} \ln\biggl(\frac{\tau}{256}\biggr)
	\nonumber \\ & \quad \quad \qquad +
	\nu\biggl(\frac{3147553127}{780337152} - \frac{451}{3072}\pi^2 \biggr)
	  - \frac{15211}{442368}\nu^2 +
	\frac{25565}{331776}\nu^3\Biggr] 
	\nonumber \\ &
	\quad \quad + \pi
	\tau^{-7/8}  \biggl(-\frac{113868647}{433520640} -
	\frac{31821}{143360}\nu + \frac{294941}{3870720}\nu^2\biggr) 
	\nonumber \\ & \quad
	\quad +  \tau^{-1}\ln\tau \Biggl[-\frac{2518977598355703073}{3779358859513036800} + \! \frac{9203}{215040}\gamma_\text{E} + \!\frac{9049}{258048}\pi^2 +\! \frac{14873}{1128960}\ln 2	+ \frac{47385}{1605632}\ln 3 - \! \frac{9203}{3440640}\ln\tau 
	\nonumber \\ & \quad \quad \qquad +
	\nu\Biggl(  \bm{\frac{696512320178557}{576825222758400}} + \frac{244493}{1128960}\gamma_\text{E} - \frac{65577}{1835008}\pi^2 + \frac{15761}{47040}\ln 2 - \frac{47385}{401408}\ln 3 - \frac{244493}{18063360}\ln\tau\Biggr)  
	\nonumber \\ & \quad \quad \qquad + \nu^2\Biggl(  - \frac{1502014727}{8323596288} + \frac{2255}{393216}\pi^2 \Biggr) - \frac{258479}{33030144} \nu^3 + \frac{1195}{262144} \nu^4 - \frac{\aleph_8}{1024}  \Biggr]
	\nonumber \\ & \quad
	\quad + \pi \,\tau^{-9/8}\Biggl[-\frac{9965202491753717}{5768252227584000} + \frac{107}{600}\gamma_\text{E} + \frac{23}{600}\pi^2 - \frac{107}{4800}\ln\biggl(\frac{\tau}{256}\biggr) 
	\nonumber \\ & \quad \quad \qquad +
	\nu\biggl( \frac{8248609881163}{2746786775040} - \frac{3157}{30720}\pi^2 \biggr)
	  - \frac{3590973803}{20808990720}\nu^2 -
	\frac{520159}{1634992128}\nu^3 \Biggr] 
	+ \mathcal{O}\bigl(\tau^{-5/4}\bigr)\Biggr\}\,.
\end{align}
\end{widetext}

We now turn to the orbital phase, which can be obtained using
\begin{align}
     \phi(y) &= \frac{c^3}{G m} \int_{0}^y \dd y' \, \frac{y'^{3/2}}{\dd y'/\dd t} \,.
\end{align}
Performing the post-Newtonian expansion of the integrand before integrating, we obtain 
\begin{widetext}
\begin{align}\label{phix}
	 \phi(y)  &= \phi_0 - \frac{y^{-5/2}}{32\nu}\Biggl\{ 1 + y\biggl(
	\frac{3715}{1008} + \frac{55}{12}\nu\biggr) - 10\pi y^{3/2}
	\nonumber \\ & \quad + y^2\biggl( \frac{15293365}{1016064} +
	\frac{27145}{1008} \nu + \frac{3085}{144} \nu^2 \biggr)  +
	\pi y^{5/2} \biggl(\frac{38645}{1344} - \frac{65}{16}\nu\biggr)  \ln y
	\nonumber \\ & \quad +
	y^3\Biggl[\frac{12348611926451}{18776862720} - \frac{160}{3}\pi^2 -
	\frac{1712}{21}\gamma_\text{E} - \frac{856}{21} \ln (16\,y)
	\nonumber \\ & \quad \qquad +
	\nu\biggl(-\frac{15737765635}{12192768} + \frac{2255}{48}\pi^2
	\biggr) + \frac{76055}{6912}\nu^2 -
	\frac{127825}{5184}\nu^3\Biggr]  
	\nonumber \\ & \quad +
	  \pi y^{7/2}\biggl(\frac{77096675}{2032128} + \frac{378515}{12096}\nu -
	\frac{74045}{6048}\nu^2\biggr)
	\nonumber \\ & \quad +
	y^{4} \Biggl[ \frac{2550713843998885153}{2214468081745920} - \frac{9203}{126}\gamma_\text{E} - \frac{45245}{756}\pi^2 - \frac{252755}{2646}\ln 2 - \frac{78975}{1568}\ln 3 - \frac{9203}{252}\ln y 
	\nonumber \\ & \quad \qquad +
	\nu\biggl(-\bm{\frac{659081900394877}{337983528960}} - \frac{488986}{1323}\gamma_\text{E} + \frac{109295}{1792}\pi^2 - \frac{1245514}{1323}\ln 2 + \frac{78975}{392}\ln 3 - \frac{244493}{1323}\ln y \biggr) 
	\nonumber \\ & \quad \qquad  +
	\nu^2\biggl( \frac{7510073635}{24385536} - \frac{11275}{1152}\pi^2 \biggr) +
	 \frac{1292395}{96768}\nu^3 - \frac{5975}{768}\nu^4 + \frac{5}{3}\aleph_8 \Biggr]  
	\nonumber \\ & \quad +
	\pi y^{9/2}  \Biggl[ - \frac{93098188434443}{150214901760} + \frac{1712}{21}\gamma_\text{E} + \frac{80}{3}\pi^2 +  \frac{856}{21}\ln (16 y)
	\nonumber \\ & \qquad \qquad +
	\nu\biggl( \frac{1492917260735}{1072963584} - \frac{2255}{48}\pi^2\biggr)
	 - \frac{45293335}{1016064}\nu^2 - \frac{10323755}{1596672}\nu^3 \Biggr] 
	 + \mathcal{O}\bigl(y^5\bigr)\Biggr\}\,.
\end{align}
\end{widetext}
In these expressions, one can ignore horizon absorption and tidal heating effects by setting $\aleph_8 =0$, or specialize to the case of two nonspinning black holes by setting $\aleph_8 = 1-4\nu +2\nu^2$.  Moreover, I have put in bold font the coefficients which differ from the analogous right-hand sides when working with \textit{waveform} frequency. The corresponding results in terms of waveform frequency are provided in the following section.

\subsection{In terms of the waveform frequency}
\label{subsec:phasing_waveform}

The previous derivation can be repeated in terms of the waveform frequency, such that $b_0$ has already dropped out from both sides of the flux balance law. This was the approach followed in \cite{BFHLTa,BFHLTb}.

The fluxes of energy and angular momentum at infinity are given in terms of the waveform frequency in Eq.~(4)~of~Ref.~\cite{BFHLTa} and Eq.~(20)~of~Ref.~\cite{Khairnar:2024rzs}, respectively. They satisfy the relation
\begin{equation} \label{eq:relation_fluxes_waveform_frequency}
    \mathcal{F}_E^\infty(x) = \Omega_{22}\, \mathcal{F}_J^\infty(x)\,.
\end{equation}
The horizon fluxes have the same expressions in terms of orbital and waveform frequency at this order, see Eqs.~\eqref{eq:EJ_flux_horizon_y}.
Using the chain rule on \eqref{eq:flux_balance_laws}, we find that the relevant formulas read
\begin{subequations}
    \begin{align}
    \frac{\dd x}{\dd t} &= - \frac{\mathcal{F}_E(x)}{\dd E(x) / \dd x}\,, \\*
    t - t_c' &= \int^{x} \frac{\dd x'}{\dd x'/dt}\,, \\*
    \psi(x) &=  \frac{c^3}{G m} \int_{0}^x \dd x' \, \frac{x'^{3/2}}{\dd x'/\dd t} \,,
\end{align}
\end{subequations}
where  $t_c'$ can differ from $t_c$ defined in the previous section.
The differential equation governing the waveform frequency  can again be linearly split into a contribution from the fluxes at infinity and another subdominant contribution from the horizon fluxes,
\begin{subequations}\begin{align}
    \frac{\dd x}{\dd t} &= \left\{\frac{\dd x}{\dd t}\right\}_{\infty} + \left\{\frac{\dd x}{\dd t}\right\}_{\mathcal{H}} \,,
\end{align}
where the two contributions are given by 
\begin{align}
    \left\{\frac{\dd x}{\dd t}\right\}_{\infty} &= \frac{c^3}{G m}\left[\frak{D}(x;\nu) - \frac{12288}{25}\nu^2 x^9\right] \,,\\*
    \left\{\frac{\dd x}{\dd t}\right\}_{\mathcal{H}} &=  \frac{64 c^3 \nu }{5 G m}   \aleph_8 \, x^{9} \,,
\end{align}\end{subequations}
and where the function $z\mapsto \frak{D}(z;\nu)$ was given in \eqref{eq:D_of_z}.
When ignoring horizon absorption and tidal dissipation, we recover exactly the results of Ref.~\cite{BFHLTa} for the waveform frequency in terms of time $x(\tau)$ and the waveform phase in terms of the waveform frequency $\psi(x)$ [see (6) and (8) in that reference].

We can then include the contributions of horizons fluxes by adding to those expressions the following corrections:
\begin{subequations}
    \begin{align}
        \delta x(\tau) &= - \frac{\aleph_8 \ln(\tau)}{4096\, \tau^{5/4}} \,,\\*
        \delta \psi(x) &= - \frac{5 \aleph_8 x^{3/2}}{96 \nu} \,.
    \end{align}
\end{subequations}

\section{Discussion}
\label{sec:discussion}

This computation of the pseudo-Schott term at 4PN has solidly confirmed the physical hypothesis used in Refs.~\cite{BFHLTa,BFHLTb} and reinforced our trust in the validity of the phasing obtained therein. 
However, the existence of this Schott term could potentially impact effective one-body models~\cite{Nagar:2024dzj,Nagar:2024oyk}, since they seem not to account for the difference between orbital and waveform frequency in their 4PN expressions. It could also shed light on the reason why analytical black hole perturbation theory needs to choose $b_0 = 2\dM/(c^2\sqrt{\de})$  to reconcile their results with post-Newtonian expressions (see Footnote 3 of Ref.~\cite{FI10}).
Moreover, this result could solve a longstanding discrepancy between self-force and post-Newtonian theory. Indeed, it was found that the Bondi energy obtained from self-force was not agreeing with the post-Newtonian conservative energy~\cite{Pound:2019lzj}, whereas the energy obtained from the first law of binary black hole mechanics or Hamiltonian methods did~\cite{LKPT25}. It was unclear at the time if an agreement with the post-Newtonian conservative energy was expected, and it was conjectured that such a difference could be caused by hereditary effects. Here, I have shown that the binding energy and conservative energy differ at 4PN, and I expect the Bondi energy from self-force to instead agree with the post-Newtonian \textit{binding} energy of Eq.~\eqref{subeq:E_binding_circ_y}.
In work to appear soon, Ref.~\cite{GLTP25} has established the precise relationship between the Bondi energy and mechanical energy in self-force theory, and has confirmed agreement with the results of this paper [provided $b_0 = 2\dM/(c^2\sqrt{\de})$],  as usual in self-force.  Future work will then investigate whether these theoretical advances are in agreement with the numerical results of \cite{Pound:2019lzj}.

One weakness of this work is the lack of a solid derivation of the postadiabatic formula obtained in Eq.~\eqref{eq:postadiabatic_integration}. I have argued that this noncausal term is introduced in some sense artificially, and that it need not converge in the future for us to define it consistently. However, a proper regularization procedure would be more satisfactory, and would be worth pursuing.

Another open question goes as follows: one could worry that the physical postulate used in Refs.~\cite{BFHLTa,BFHLTb} cannot hold at higher PN orders, because there is no reason to select the waveform frequency to be that of the $(2,2)$ mode rather than any other $(\ell,m)$ mode. However, one can hypothesize that it is possible to define an effective frequency $\Omega_\text{eff}$ which would generalize such an observation. Namely, if the PN expansion of the conservative energy and angular momentum in terms of the orbital frequency read 
\begin{subequations}
    \begin{align}
    E_\text{cons}^\text{circ}(y) &= \sum_{n,k}  \alpha_n(\nu) \, y^{n/2} (\ln y)^k\,, \\*
    J_\text{cons}^\text{circ}(y) &= \sum_{n,k}  \beta_n(\nu)\, y^{n/2} (\ln y)^k \,,
\end{align}
\end{subequations}
then we can hypothesize that the binding energy and angular momentum (including Schott terms) read in terms of the effective frequency
such that
\begin{subequations}
\begin{align}
    E^\text{circ}(x_\text{eff}) &\overset{?}{=} \sum_{n,k}  \alpha_n(\nu) x_\text{eff}^{n/2} (\ln x_\text{eff})^k\,,\\*
    J^\text{circ}(x_\text{eff}) &\overset{?}{=} \sum_{n,k}  \beta_n(\nu) x_\text{eff}^{n/2} (\ln x_\text{eff})^k\,,
\end{align}
\end{subequations}
where $x_\text{eff} = \left({G m \,\Omega_\text{eff}}/{c^3}\right) ^{2/3}$. A consequence would be that the binding energy and angular momentum satisfy
\begin{align}
    \frac{\dd E^\text{circ}}{\dd\Omega_\text{eff}} \overset{?}{=} \Omega_\text{eff}\, \frac{\dd J^\text{circ}}{\dd \Omega_\text{eff}} \,,
\end{align}
which can be rephrased as the first law of binary black hole mechanics (with masses held fixed) \cite{LBW12} in terms of this effective frequency
\begin{align}\label{eq:first_law_omega_eff}
    \delta E \overset{?}{=} \Omega_\text{eff}\,\delta J \,.
\end{align}
%

Now, how should we define $\Omega_\text{eff}$? At 4PN, it should correspond to $\Omega_{22}$, because the $(2,2)$ is the dominant contribution to the gravitational radiation. A natural definition for $\Omega_\text{eff}$ would be a linear combination of every $(\ell,m)$ frequency weighted by their relative contribution to the gravitational radiation at infinity, namely 
\begin{align}\label{eq:omega_eff_expression}
    \Omega_\text{eff} \overset{?}{=} \sum_{\ell m} \frac{\mathcal{F}_{\ell m}}{\mathcal{F}} \Omega_{\ell m} \,,
\end{align}
where $\mathcal{F}$ can be either the energy and angular momentum flux, and $\mathcal{F}_{\ell m}$ the contribution from the $h_{\ell m}$ mode (see \cite{Warburton:2024xnr} for explicit expressions). Naturally, $\Omega_\text{eff}$ coincides with the orbital frequency up to 3.5PN order, with $\Omega_{22}$ up to 4.5PN, and becomes nontrivial beyond 5PN. 

As a matter of fact,  preliminary work in Ref.~\cite{GLTP25} has shown that the proposed $\Omega_\text{eff}$ of Eq.~\eqref{eq:omega_eff_expression} succeeds in extending the first law of binary black holes mechanics as in Eq.~\eqref{eq:first_law_omega_eff} to all PN order and at second order in the mass ratio, with masses being kept fixed and horizon effects ignored. One could think of extensions to include these horizon effects and extend the argument to all orders in the mass ratio.

\acknowledgments

This work heavily builds on results obtained in collaboration with Luc~Blanchet and Guillaume~Faye~\mbox{\cite{TB23_ToM,BFHLTa, BFHLTb, BFT24}}, who have also taught me many of the techniques used in this paper.
I would also like to thank Adam Pound for many fruitful discussions about the slicing (in)dependence of the flux-balance laws, Thibault Damour for suggesting to me the possible relevance of a Schott term, Alexandre Le Tiec for carefully proof-reading the manuscript and useful comments, as well as Alexander M. Grant for interesting discussions. I acknowledge support from the ERC Consolidator/UKRI Frontier Research Grant GWModels (selected by the ERC and funded by UKRI [Grant No. EP/Y008251/1]).
\appendix
\begin{widetext}
\section{Quadrupole moment at 2.5PN for circular orbits}
\label{app:quadrupole}

In the center of mass frame and for circular orbits, the successive time derivatives of the quadrupole moment are given by
    \begin{subequations}\begin{align}
        \dM_{ij} &= \frac{G^2 m^3 \nu}{c^4 y^2} \Bigg\{n^{\langle i}n^{j \rangle} \Bigg[1+y\left(- \frac{85}{42}- \frac{11}{42}\nu \right) + y^2\left(\frac{1087}{1512} - \frac{512}{216}\nu - \frac{205}{1512}\nu^2\right)\Bigg] \nonumber\\
        &\qquad\qquad\,\, + \lambda^{\langle i}\lambda^{j \rangle} \Bigg[y\left(\frac{11}{12} - \frac{11}{7}\nu\right) + y^2 \left(\frac{815}{378}+ \frac{137}{54}\nu - \frac{563}{378}\nu^2\right)\Bigg]  + \frac{48}{7} \nu y^{5/2} n^{\langle i}\lambda^{j \rangle} \Bigg\}\,,\\
        \dM^{(1)}_{ij} &=  \frac{2 G m^2 \nu}{c \sqrt{y}} \Bigg\{n^{\langle i}\lambda^{j \rangle} \Bigg[1+y\left(- \frac{107}{42} + \frac{55}{42}\nu \right) + y^2\left(-\frac{2173}{1512} - \frac{1069}{216}\nu + \frac{2047}{1512}\nu^2\right)\Bigg] \nonumber\\
        &\qquad\qquad\,\, - \frac{568}{35} \nu y^{5/2} n^{\langle i}n^{j \rangle} + \frac{24}{7}\nu y^{5/2}\lambda^{\langle i}\lambda^{j \rangle}    \Bigg\}\\
        \dM^{(2)}_{ij} &= -2 m \nu c^2 y  \Bigg\{\left(n^{\langle i}n^{j \rangle} - \lambda^{\langle i}\lambda^{j \rangle}\right)\Bigg[1+y\left(- \frac{107}{42}+ \frac{55}{42}\nu \right) + y^2\left(-\frac{2173}{1512} - \frac{1069}{216}\nu + \frac{2047}{1512}\nu^2\right)\Bigg] \nonumber\\
        &\qquad\qquad\,\,     + \frac{320}{7} \nu y^{5/2} n^{\langle i}\lambda^{j \rangle} \Bigg\} \\
        \dM^{(3)}_{ij} &= - \frac{8\nu  c^5 y^{5/2} }{G}\Bigg\{ n^{\langle i}\lambda^{j \rangle} \Bigg[1+y\left(- \frac{107}{42} + \frac{55}{42}\nu \right) + y^2\left(-\frac{2173}{1512} - \frac{1069}{216}\nu + \frac{2047}{1512}\nu^2\right)\Bigg]\nonumber\\
        &\qquad\qquad\,\, - \frac{288}{35} \nu y^{5/2} \left(n^{\langle i}n^{j \rangle} - \lambda^{\langle i}\lambda^{j \rangle}\right)  \Bigg\}\\
        \dM^{(4)}_{ij} &= \frac{8 \nu c^8 y^4}{G^2 m}\Bigg\{\left(n^{\langle i}n^{j \rangle} - \lambda^{\langle i}\lambda^{j \rangle}\right)\Bigg[1+y\left(- \frac{107}{42}+ \frac{55}{42}\nu \right) + y^2\left(-\frac{2173}{1512} - \frac{1069}{216}\nu + \frac{2047}{1512}\nu^2\right)\Bigg] \nonumber\\
        &\qquad\qquad\,\,     + \frac{32}{35} \nu y^{5/2} n^{\langle i}\lambda^{j \rangle}  \Bigg\}\\
        \dM^{(5)}_{ij} &= \frac{32 \nu c^{11} y^{11/2}}{G^3 m^2}\Bigg\{ n^{\langle i}\lambda^{j \rangle} \Bigg[1+y\left(- \frac{107}{42} + \frac{55}{42}\nu \right) + y^2\left(-\frac{2173}{1512} - \frac{1069}{216}\nu + \frac{2047}{1512}\nu^2\right)\Bigg]\nonumber\\
        &\qquad\qquad\,\, + \frac{88}{7} \nu y^{5/2} \left(n^{\langle i}n^{j \rangle} - \lambda^{\langle i}\lambda^{j \rangle}\right)  \Bigg\}\\
        \dM^{(6)}_{ij} &= -\frac{32 \nu c^{14} y^{7}}{G^4 m^3}\Bigg\{\left(n^{\langle i}n^{j \rangle} - \lambda^{\langle i}\lambda^{j \rangle}\right)\Bigg[1+y\left(- \frac{107}{42}+ \frac{55}{42}\nu \right) + y^2\left(-\frac{2173}{1512} - \frac{1069}{216}\nu + \frac{2047}{1512}\nu^2\right)\Bigg] \nonumber\\
        &\qquad\qquad\,\,  - \frac{4224}{35} \nu y^{5/2} n^{\langle i}\lambda^{j \rangle}  \Bigg\} \\
        \dM^{(7)}_{ij} &= -\frac{128 \nu c^{17} y^{17/2}}{G^5 m^4}\Bigg\{ n^{\langle i}\lambda^{j \rangle} \Bigg[1+y\left(- \frac{107}{42} + \frac{55}{42}\nu \right) + y^2\left(-\frac{2173}{1512} - \frac{1069}{216}\nu + \frac{2047}{1512}\nu^2\right)\Bigg]\nonumber\\
        &\qquad\qquad\,\, + \frac{368}{7} \nu y^{5/2} \left(n^{\langle i}n^{j \rangle} - \lambda^{\langle i}\lambda^{j \rangle}\right)  \Bigg\}
    \end{align}\end{subequations}
\end{widetext}
~
\pagebreak
\bibliography{references}

\begin{thebibliography}{70}%
\makeatletter
\providecommand \@ifxundefined [1]{%
 \@ifx{#1\undefined}
}%
\providecommand \@ifnum [1]{%
 \ifnum #1\expandafter \@firstoftwo
 \else \expandafter \@secondoftwo
 \fi
}%
\providecommand \@ifx [1]{%
 \ifx #1\expandafter \@firstoftwo
 \else \expandafter \@secondoftwo
 \fi
}%
\providecommand \natexlab [1]{#1}%
\providecommand \enquote  [1]{``#1''}%
\providecommand \bibnamefont  [1]{#1}%
\providecommand \bibfnamefont [1]{#1}%
\providecommand \citenamefont [1]{#1}%
\providecommand \href@noop [0]{\@secondoftwo}%
\providecommand \href [0]{\begingroup \@sanitize@url \@href}%
\providecommand \@href[1]{\@@startlink{#1}\@@href}%
\providecommand \@@href[1]{\endgroup#1\@@endlink}%
\providecommand \@sanitize@url [0]{\catcode `\\12\catcode `\$12\catcode
  `\&12\catcode `\#12\catcode `\^12\catcode `\_12\catcode `\%12\relax}%
\providecommand \@@startlink[1]{}%
\providecommand \@@endlink[0]{}%
\providecommand \url  [0]{\begingroup\@sanitize@url \@url }%
\providecommand \@url [1]{\endgroup\@href {#1}{\urlprefix }}%
\providecommand \urlprefix  [0]{URL }%
\providecommand \Eprint [0]{\href }%
\providecommand \doibase [0]{https://doi.org/}%
\providecommand \selectlanguage [0]{\@gobble}%
\providecommand \bibinfo  [0]{\@secondoftwo}%
\providecommand \bibfield  [0]{\@secondoftwo}%
\providecommand \translation [1]{[#1]}%
\providecommand \BibitemOpen [0]{}%
\providecommand \bibitemStop [0]{}%
\providecommand \bibitemNoStop [0]{.\EOS\space}%
\providecommand \EOS [0]{\spacefactor3000\relax}%
\providecommand \BibitemShut  [1]{\csname bibitem#1\endcsname}%
\let\auto@bib@innerbib\@empty
\bibitem [{\citenamefont {Bini}\ and\ \citenamefont {Damour}(2013)}]{BiniD13}%
  \BibitemOpen
  \bibfield  {author} {\bibinfo {author} {\bibfnamefont {D.}~\bibnamefont
  {Bini}}\ and\ \bibinfo {author} {\bibfnamefont {T.}~\bibnamefont {Damour}},\
  }\bibfield  {title} {\bibinfo {title} {{Analytical determination of the
  two-body gravitational interaction potential at the fourth post-Newtonian
  approximation}},\ }\href {https://doi.org/10.1103/PhysRevD.87.121501}
  {\bibfield  {journal} {\bibinfo  {journal} {Phys. Rev. D}\ }\textbf {\bibinfo
  {volume} {87}},\ \bibinfo {pages} {121501(R)} (\bibinfo {year} {2013})},\
  \Eprint {https://arxiv.org/abs/1305.4884} {arXiv:1305.4884 [gr-qc]}
  \BibitemShut {NoStop}%
\bibitem [{\citenamefont {Damour}\ \emph
  {et~al.}(2014{\natexlab{a}})\citenamefont {Damour}, \citenamefont
  {Jaranowski},\ and\ \citenamefont {Sch{\"a}fer}}]{DJS14}%
  \BibitemOpen
  \bibfield  {author} {\bibinfo {author} {\bibfnamefont {T.}~\bibnamefont
  {Damour}}, \bibinfo {author} {\bibfnamefont {P.}~\bibnamefont {Jaranowski}},\
  and\ \bibinfo {author} {\bibfnamefont {G.}~\bibnamefont {Sch{\"a}fer}},\
  }\bibfield  {title} {\bibinfo {title} {{Non-local-in-time action for the
  fourth post-Newtonian conservative dynamics of two-body systems}},\
  }\href@noop {} {\bibfield  {journal} {\bibinfo  {journal} {Phys. Rev. D}\
  }\textbf {\bibinfo {volume} {89}},\ \bibinfo {pages} {064058} (\bibinfo
  {year} {2014}{\natexlab{a}})},\ \Eprint {https://arxiv.org/abs/1401.4548}
  {arXiv:1401.4548 [gr-qc]} \BibitemShut {NoStop}%
\bibitem [{\citenamefont {Damour}\ \emph
  {et~al.}(2014{\natexlab{b}})\citenamefont {Damour}, \citenamefont
  {Jaranowski},\ and\ \citenamefont {Sch{\"a}fer}}]{Damour:2014jta}%
  \BibitemOpen
  \bibfield  {author} {\bibinfo {author} {\bibfnamefont {T.}~\bibnamefont
  {Damour}}, \bibinfo {author} {\bibfnamefont {P.}~\bibnamefont {Jaranowski}},\
  and\ \bibinfo {author} {\bibfnamefont {G.}~\bibnamefont {Sch{\"a}fer}},\
  }\bibfield  {title} {\bibinfo {title} {{Nonlocal-in-time action for the
  fourth post-Newtonian conservative dynamics of two-body systems}},\ }\href
  {https://doi.org/10.1103/PhysRevD.89.064058} {\bibfield  {journal} {\bibinfo
  {journal} {Phys. Rev. D}\ }\textbf {\bibinfo {volume} {89}},\ \bibinfo
  {pages} {064058} (\bibinfo {year} {2014}{\natexlab{b}})},\ \Eprint
  {https://arxiv.org/abs/1401.4548} {arXiv:1401.4548 [gr-qc]} \BibitemShut
  {NoStop}%
\bibitem [{\citenamefont {Jaranowski}\ and\ \citenamefont
  {Sch{\"a}fer}(2015)}]{JaraS15}%
  \BibitemOpen
  \bibfield  {author} {\bibinfo {author} {\bibfnamefont {P.}~\bibnamefont
  {Jaranowski}}\ and\ \bibinfo {author} {\bibfnamefont {G.}~\bibnamefont
  {Sch{\"a}fer}},\ }\bibfield  {title} {\bibinfo {title} {{Derivation of the
  local-in-time fourth post-Newtonian ADM Hamiltonian for spinless compact
  binaries}},\ }\href@noop {} {\bibfield  {journal} {\bibinfo  {journal} {Phys.
  Rev. D}\ }\textbf {\bibinfo {volume} {92}},\ \bibinfo {pages} {124043}
  (\bibinfo {year} {2015})},\ \Eprint {https://arxiv.org/abs/1508.01016}
  {arXiv:1508.01016 [gr-qc]} \BibitemShut {NoStop}%
\bibitem [{\citenamefont {Damour}\ \emph {et~al.}(2015)\citenamefont {Damour},
  \citenamefont {Jaranowski},\ and\ \citenamefont {Sch{\"a}fer}}]{DJS15eob}%
  \BibitemOpen
  \bibfield  {author} {\bibinfo {author} {\bibfnamefont {T.}~\bibnamefont
  {Damour}}, \bibinfo {author} {\bibfnamefont {P.}~\bibnamefont {Jaranowski}},\
  and\ \bibinfo {author} {\bibfnamefont {G.}~\bibnamefont {Sch{\"a}fer}},\
  }\bibfield  {title} {\bibinfo {title} {{Fourth post-Newtonian effective
  one-body dynamics}},\ }\href@noop {} {\bibfield  {journal} {\bibinfo
  {journal} {Phys. Rev. D}\ }\textbf {\bibinfo {volume} {91}},\ \bibinfo
  {pages} {084024} (\bibinfo {year} {2015})},\ \Eprint
  {https://arxiv.org/abs/1502.07245} {arXiv:1502.07245 [gr-qc]} \BibitemShut
  {NoStop}%
\bibitem [{\citenamefont {Damour}\ \emph {et~al.}(2016)\citenamefont {Damour},
  \citenamefont {Jaranowski},\ and\ \citenamefont {Sch{\"a}fer}}]{DJS16}%
  \BibitemOpen
  \bibfield  {author} {\bibinfo {author} {\bibfnamefont {T.}~\bibnamefont
  {Damour}}, \bibinfo {author} {\bibfnamefont {P.}~\bibnamefont {Jaranowski}},\
  and\ \bibinfo {author} {\bibfnamefont {G.}~\bibnamefont {Sch{\"a}fer}},\
  }\bibfield  {title} {\bibinfo {title} {{On the conservative dynamics of
  two-body systems at the fourth post-Newtonian approximation of general
  relativity}},\ }\href {https://doi.org/10.1103/PhysRevD.93.084014} {\bibfield
   {journal} {\bibinfo  {journal} {Phys. Rev. D}\ }\textbf {\bibinfo {volume}
  {93}},\ \bibinfo {pages} {084014} (\bibinfo {year} {2016})},\ \Eprint
  {https://arxiv.org/abs/1601.01283} {arXiv:1601.01283 [gr-qc]} \BibitemShut
  {NoStop}%
\bibitem [{\citenamefont {Bernard}\ \emph {et~al.}(2016)\citenamefont
  {Bernard}, \citenamefont {Blanchet}, \citenamefont {Boh\'e}, \citenamefont
  {Faye},\ and\ \citenamefont {Marsat}}]{BBBFMa}%
  \BibitemOpen
  \bibfield  {author} {\bibinfo {author} {\bibfnamefont {L.}~\bibnamefont
  {Bernard}}, \bibinfo {author} {\bibfnamefont {L.}~\bibnamefont {Blanchet}},
  \bibinfo {author} {\bibfnamefont {A.}~\bibnamefont {Boh\'e}}, \bibinfo
  {author} {\bibfnamefont {G.}~\bibnamefont {Faye}},\ and\ \bibinfo {author}
  {\bibfnamefont {S.}~\bibnamefont {Marsat}},\ }\bibfield  {title} {\bibinfo
  {title} {{Fokker action of nonspinning compact binaries at the fourth
  post-Newtonian approximation}},\ }\href
  {https://doi.org/10.1103/PhysRevD.93.084037} {\bibfield  {journal} {\bibinfo
  {journal} {Phys. Rev. D}\ }\textbf {\bibinfo {volume} {93}},\ \bibinfo
  {pages} {084037} (\bibinfo {year} {2016})},\ \Eprint
  {https://arxiv.org/abs/1512.02876} {arXiv:1512.02876 [gr-qc]} \BibitemShut
  {NoStop}%
\bibitem [{\citenamefont {Bernard}\ \emph
  {et~al.}(2017{\natexlab{a}})\citenamefont {Bernard}, \citenamefont
  {Blanchet}, \citenamefont {Boh\'e}, \citenamefont {Faye},\ and\ \citenamefont
  {Marsat}}]{BBBFMb}%
  \BibitemOpen
  \bibfield  {author} {\bibinfo {author} {\bibfnamefont {L.}~\bibnamefont
  {Bernard}}, \bibinfo {author} {\bibfnamefont {L.}~\bibnamefont {Blanchet}},
  \bibinfo {author} {\bibfnamefont {A.}~\bibnamefont {Boh\'e}}, \bibinfo
  {author} {\bibfnamefont {G.}~\bibnamefont {Faye}},\ and\ \bibinfo {author}
  {\bibfnamefont {S.}~\bibnamefont {Marsat}},\ }\bibfield  {title} {\bibinfo
  {title} {{Energy and periastron advance of compact binaries on circular
  orbits at the fourth post-Newtonian order}},\ }\href
  {https://doi.org/10.1103/PhysRevD.95.044026} {\bibfield  {journal} {\bibinfo
  {journal} {Phys. Rev. D}\ }\textbf {\bibinfo {volume} {95}},\ \bibinfo
  {pages} {044026} (\bibinfo {year} {2017}{\natexlab{a}})},\ \Eprint
  {https://arxiv.org/abs/1610.07934} {arXiv:1610.07934 [gr-qc]} \BibitemShut
  {NoStop}%
\bibitem [{\citenamefont {Bernard}\ \emph
  {et~al.}(2017{\natexlab{b}})\citenamefont {Bernard}, \citenamefont
  {Blanchet}, \citenamefont {Boh\'e}, \citenamefont {Faye},\ and\ \citenamefont
  {Marsat}}]{BBBFMc}%
  \BibitemOpen
  \bibfield  {author} {\bibinfo {author} {\bibfnamefont {L.}~\bibnamefont
  {Bernard}}, \bibinfo {author} {\bibfnamefont {L.}~\bibnamefont {Blanchet}},
  \bibinfo {author} {\bibfnamefont {A.}~\bibnamefont {Boh\'e}}, \bibinfo
  {author} {\bibfnamefont {G.}~\bibnamefont {Faye}},\ and\ \bibinfo {author}
  {\bibfnamefont {S.}~\bibnamefont {Marsat}},\ }\bibfield  {title} {\bibinfo
  {title} {{Dimensional regularization of the IR divergences in the Fokker
  action of point-particle binaries at the fourth post-Newtonian order}},\
  }\href {https://doi.org/10.1103/PhysRevD.96.104043} {\bibfield  {journal}
  {\bibinfo  {journal} {Phys. Rev. D}\ }\textbf {\bibinfo {volume} {96}},\
  \bibinfo {pages} {104043} (\bibinfo {year} {2017}{\natexlab{b}})},\ \Eprint
  {https://arxiv.org/abs/1706.08480} {arXiv:1706.08480 [gr-qc]} \BibitemShut
  {NoStop}%
\bibitem [{\citenamefont {Bernard}\ \emph {et~al.}(2018)\citenamefont
  {Bernard}, \citenamefont {Blanchet}, \citenamefont {Faye},\ and\
  \citenamefont {Marchand}}]{BBFM17}%
  \BibitemOpen
  \bibfield  {author} {\bibinfo {author} {\bibfnamefont {L.}~\bibnamefont
  {Bernard}}, \bibinfo {author} {\bibfnamefont {L.}~\bibnamefont {Blanchet}},
  \bibinfo {author} {\bibfnamefont {G.}~\bibnamefont {Faye}},\ and\ \bibinfo
  {author} {\bibfnamefont {T.}~\bibnamefont {Marchand}},\ }\bibfield  {title}
  {\bibinfo {title} {{Center-of-Mass Equations of Motion and Conserved
  Integrals of Compact Binary Systems at the Fourth Post-Newtonian Order}},\
  }\href {https://doi.org/10.1103/PhysRevD.97.044037} {\bibfield  {journal}
  {\bibinfo  {journal} {Phys. Rev. D}\ }\textbf {\bibinfo {volume} {97}},\
  \bibinfo {pages} {044037} (\bibinfo {year} {2018})},\ \Eprint
  {https://arxiv.org/abs/1711.00283} {arXiv:1711.00283 [gr-qc]} \BibitemShut
  {NoStop}%
\bibitem [{\citenamefont {Foffa}\ and\ \citenamefont {Sturani}(2012)}]{FStail}%
  \BibitemOpen
  \bibfield  {author} {\bibinfo {author} {\bibfnamefont {S.}~\bibnamefont
  {Foffa}}\ and\ \bibinfo {author} {\bibfnamefont {R.}~\bibnamefont
  {Sturani}},\ }\bibfield  {title} {\bibinfo {title} {{Tail terms in
  gravitational radiation reaction via effective field theory}},\ }\href
  {https://doi.org/10.1103/PhysRevD.87.044056} {\bibfield  {journal} {\bibinfo
  {journal} {Phys. Rev. D}\ }\textbf {\bibinfo {volume} {87}},\ \bibinfo
  {pages} {044056} (\bibinfo {year} {2012})},\ \Eprint
  {https://arxiv.org/abs/1111.5488} {arXiv:1111.5488 [gr-qc]} \BibitemShut
  {NoStop}%
\bibitem [{\citenamefont {Foffa}\ and\ \citenamefont {Sturani}(2013)}]{FS4PN}%
  \BibitemOpen
  \bibfield  {author} {\bibinfo {author} {\bibfnamefont {S.}~\bibnamefont
  {Foffa}}\ and\ \bibinfo {author} {\bibfnamefont {R.}~\bibnamefont
  {Sturani}},\ }\bibfield  {title} {\bibinfo {title} {{The dynamics of the
  gravitational two-body problem in the post-Newtonian approximation at
  quadratic order in the Newton's constant}},\ }\href
  {https://doi.org/10.1103/PhysRevD.87.064011} {\bibfield  {journal} {\bibinfo
  {journal} {Phys. Rev. D}\ }\textbf {\bibinfo {volume} {87}},\ \bibinfo
  {pages} {064011} (\bibinfo {year} {2013})},\ \Eprint
  {https://arxiv.org/abs/1206.7087} {arXiv:1206.7087 [gr-qc]} \BibitemShut
  {NoStop}%
\bibitem [{\citenamefont {Foffa}\ \emph {et~al.}(2017)\citenamefont {Foffa},
  \citenamefont {Mastrolia}, \citenamefont {Sturani},\ and\ \citenamefont
  {Sturm}}]{FMSS16}%
  \BibitemOpen
  \bibfield  {author} {\bibinfo {author} {\bibfnamefont {S.}~\bibnamefont
  {Foffa}}, \bibinfo {author} {\bibfnamefont {P.}~\bibnamefont {Mastrolia}},
  \bibinfo {author} {\bibfnamefont {R.}~\bibnamefont {Sturani}},\ and\ \bibinfo
  {author} {\bibfnamefont {C.}~\bibnamefont {Sturm}},\ }\bibfield  {title}
  {\bibinfo {title} {{Effective field theory approach to the gravitational
  two-body dynamics at fourth post-Newtonian order and quintic in the Newton
  constant}},\ }\href {https://doi.org/10.1103/PhysRevD.95.104009} {\bibfield
  {journal} {\bibinfo  {journal} {Phys. Rev. D}\ }\textbf {\bibinfo {volume}
  {95}},\ \bibinfo {pages} {104009} (\bibinfo {year} {2017})},\ \Eprint
  {https://arxiv.org/abs/1612.00482} {arXiv:1612.00482 [gr-qc]} \BibitemShut
  {NoStop}%
\bibitem [{\citenamefont {Foffa}\ and\ \citenamefont {Sturani}(2019)}]{FS19}%
  \BibitemOpen
  \bibfield  {author} {\bibinfo {author} {\bibfnamefont {S.}~\bibnamefont
  {Foffa}}\ and\ \bibinfo {author} {\bibfnamefont {R.}~\bibnamefont
  {Sturani}},\ }\bibfield  {title} {\bibinfo {title} {{Conservative dynamics of
  binary systems to fourth Post-Newtonian order in the EFT approach I:
  Regularized Lagrangian}},\ }\href
  {https://doi.org/doi.org/10.1103/PhysRevD.100.024047} {\bibfield  {journal}
  {\bibinfo  {journal} {Phys. Rev. D}\ }\textbf {\bibinfo {volume} {100}},\
  \bibinfo {pages} {024047} (\bibinfo {year} {2019})},\ \Eprint
  {https://arxiv.org/abs/1903.05113} {arXiv:1903.05113 [gr-qc]} \BibitemShut
  {NoStop}%
\bibitem [{\citenamefont {Foffa}\ \emph {et~al.}(2019)\citenamefont {Foffa},
  \citenamefont {Porto}, \citenamefont {Rothstein},\ and\ \citenamefont
  {Sturani}}]{FPRS19}%
  \BibitemOpen
  \bibfield  {author} {\bibinfo {author} {\bibfnamefont {S.}~\bibnamefont
  {Foffa}}, \bibinfo {author} {\bibfnamefont {R.}~\bibnamefont {Porto}},
  \bibinfo {author} {\bibfnamefont {I.}~\bibnamefont {Rothstein}},\ and\
  \bibinfo {author} {\bibfnamefont {R.}~\bibnamefont {Sturani}},\ }\bibfield
  {title} {\bibinfo {title} {{Conservative dynamics of binary systems to fourth
  Post-Newtonian order in the EFT approach II: Renormalized Lagrangian}},\
  }\href@noop {} {\bibfield  {journal} {\bibinfo  {journal} {Phys. Rev. D}\
  }\textbf {\bibinfo {volume} {100}},\ \bibinfo {pages} {024048} (\bibinfo
  {year} {2019})},\ \Eprint {https://arxiv.org/abs/1903.05118}
  {arXiv:1903.05118 [gr-qc]} \BibitemShut {NoStop}%
\bibitem [{\citenamefont {Bl\"umlein}\ \emph {et~al.}(2020)\citenamefont
  {Bl\"umlein}, \citenamefont {Maier}, \citenamefont {Marquard},\ and\
  \citenamefont {Sch\"afer}}]{Blumlein20}%
  \BibitemOpen
  \bibfield  {author} {\bibinfo {author} {\bibfnamefont {J.}~\bibnamefont
  {Bl\"umlein}}, \bibinfo {author} {\bibfnamefont {A.}~\bibnamefont {Maier}},
  \bibinfo {author} {\bibfnamefont {P.}~\bibnamefont {Marquard}},\ and\
  \bibinfo {author} {\bibfnamefont {G.}~\bibnamefont {Sch\"afer}},\ }\bibfield
  {title} {\bibinfo {title} {{Fourth post-Newtonian Hamiltonian dynamics of
  two-body systems from an effective field theory approach}},\ }\href
  {https://doi.org/10.1016/j.nuclphysb.2020.115041} {\bibfield  {journal}
  {\bibinfo  {journal} {Nucl. Phys. B}\ }\textbf {\bibinfo {volume} {955}},\
  \bibinfo {pages} {115041} (\bibinfo {year} {2020})},\ \Eprint
  {https://arxiv.org/abs/2003.01692} {arXiv:2003.01692 [gr-qc]} \BibitemShut
  {NoStop}%
\bibitem [{\citenamefont {Burke}\ and\ \citenamefont {Thorne}(1970)}]{BuTh70}%
  \BibitemOpen
  \bibfield  {author} {\bibinfo {author} {\bibfnamefont {W.}~\bibnamefont
  {Burke}}\ and\ \bibinfo {author} {\bibfnamefont {K.}~\bibnamefont {Thorne}},\
  }\bibfield  {title} {\bibinfo {title} {{Gravitational Radiation Damping}},\
  }in\ \href@noop {} {\emph {\bibinfo {booktitle} {Relativity}}},\ \bibinfo
  {editor} {edited by\ \bibinfo {editor} {\bibfnamefont {M.}~\bibnamefont
  {Carmeli}}, \bibinfo {editor} {\bibfnamefont {S.}~\bibnamefont {Fickler}},\
  and\ \bibinfo {editor} {\bibfnamefont {L.}~\bibnamefont {Witten}}}\ (\bibinfo
   {publisher} {Plenum Press},\ \bibinfo {address} {New York and London},\
  \bibinfo {year} {1970})\ pp.\ \bibinfo {pages} {209--228}\BibitemShut
  {NoStop}%
\bibitem [{\citenamefont {Iyer}\ and\ \citenamefont {Will}(1993)}]{IW93}%
  \BibitemOpen
  \bibfield  {author} {\bibinfo {author} {\bibfnamefont {B.}~\bibnamefont
  {Iyer}}\ and\ \bibinfo {author} {\bibfnamefont {C.}~\bibnamefont {Will}},\
  }\bibfield  {title} {\bibinfo {title} {{Post-Newtonian
  gravitational-radiation reaction for two-body systems}},\ }\href@noop {}
  {\bibfield  {journal} {\bibinfo  {journal} {Phys. Rev. Lett.}\ }\textbf
  {\bibinfo {volume} {70}},\ \bibinfo {pages} {113} (\bibinfo {year}
  {1993})}\BibitemShut {NoStop}%
\bibitem [{\citenamefont {Iyer}\ and\ \citenamefont {Will}(1995)}]{IW95}%
  \BibitemOpen
  \bibfield  {author} {\bibinfo {author} {\bibfnamefont {B.}~\bibnamefont
  {Iyer}}\ and\ \bibinfo {author} {\bibfnamefont {C.}~\bibnamefont {Will}},\
  }\bibfield  {title} {\bibinfo {title} {{Post-Newtonian gravitational
  radiation reaction for two-body systems: Nonspinning bodies}},\ }\href@noop
  {} {\bibfield  {journal} {\bibinfo  {journal} {Phys. Rev. D}\ }\textbf
  {\bibinfo {volume} {52}},\ \bibinfo {pages} {6882} (\bibinfo {year}
  {1995})}\BibitemShut {NoStop}%
\bibitem [{\citenamefont {Pati}\ and\ \citenamefont {Will}(2002)}]{PW02}%
  \BibitemOpen
  \bibfield  {author} {\bibinfo {author} {\bibfnamefont {M.}~\bibnamefont
  {Pati}}\ and\ \bibinfo {author} {\bibfnamefont {C.}~\bibnamefont {Will}},\
  }\bibfield  {title} {\bibinfo {title} {{Post-Newtonian gravitational
  radiation and equations of motion via direct integration of the relaxed
  Einstein equations. II. Two-body equations of motion to second post-Newtonian
  order, and radiation-reaction to 3.5 post-Newtonian order}},\ }\href
  {https://doi.org/10.1103/PhysRevD.65.104008} {\bibfield  {journal} {\bibinfo
  {journal} {Phys. Rev. D}\ }\textbf {\bibinfo {volume} {65}},\ \bibinfo
  {pages} {104008} (\bibinfo {year} {2002})},\ \Eprint
  {https://arxiv.org/abs/gr-qc/0201001} {arXiv:gr-qc/0201001} \BibitemShut
  {NoStop}%
\bibitem [{\citenamefont {K{\"o}nigsd{\"o}rffer}\ \emph
  {et~al.}(2003)\citenamefont {K{\"o}nigsd{\"o}rffer}, \citenamefont {Faye},\
  and\ \citenamefont {Sch{\"a}fer}}]{KFS03}%
  \BibitemOpen
  \bibfield  {author} {\bibinfo {author} {\bibfnamefont {C.}~\bibnamefont
  {K{\"o}nigsd{\"o}rffer}}, \bibinfo {author} {\bibfnamefont {G.}~\bibnamefont
  {Faye}},\ and\ \bibinfo {author} {\bibfnamefont {G.}~\bibnamefont
  {Sch{\"a}fer}},\ }\bibfield  {title} {\bibinfo {title} {{The binary
  black-hole dynamics at the third-and-a-half post-Newtonian order in the
  ADM-formalism}},\ }\href {https://doi.org/10.1086/376598} {\bibfield
  {journal} {\bibinfo  {journal} {Phys. Rev. D}\ }\textbf {\bibinfo {volume}
  {68}},\ \bibinfo {pages} {044004} (\bibinfo {year} {2003})},\ \Eprint
  {https://arxiv.org/abs/astro-ph/0305048} {astro-ph/0305048} \BibitemShut
  {NoStop}%
\bibitem [{\citenamefont {Nissanke}\ and\ \citenamefont
  {Blanchet}(2005)}]{NB05}%
  \BibitemOpen
  \bibfield  {author} {\bibinfo {author} {\bibfnamefont {S.}~\bibnamefont
  {Nissanke}}\ and\ \bibinfo {author} {\bibfnamefont {L.}~\bibnamefont
  {Blanchet}},\ }\bibfield  {title} {\bibinfo {title} {{Gravitational radiation
  reaction in the equations of motion of compact binaries to 3.5 post-Newtonian
  order}},\ }\href {https://doi.org/10.1088/0264-9381/22/6/008} {\bibfield
  {journal} {\bibinfo  {journal} {Class. Quant. Grav.}\ }\textbf {\bibinfo
  {volume} {22}},\ \bibinfo {pages} {1007} (\bibinfo {year} {2005})},\ \Eprint
  {https://arxiv.org/abs/gr-qc/0412018} {arXiv:gr-qc/0412018} \BibitemShut
  {NoStop}%
\bibitem [{\citenamefont {Itoh}(2009)}]{itoh3}%
  \BibitemOpen
  \bibfield  {author} {\bibinfo {author} {\bibfnamefont {Y.}~\bibnamefont
  {Itoh}},\ }\bibfield  {title} {\bibinfo {title} {{Third-and-a-half order
  post-Newtonian equations of motion for relativistic compact binaries using
  the strong field point particle limit}},\ }\href
  {https://doi.org/10.1103/PhysRevD.80.124003} {\bibfield  {journal} {\bibinfo
  {journal} {Phys. Rev. D}\ }\textbf {\bibinfo {volume} {80}},\ \bibinfo
  {pages} {024003} (\bibinfo {year} {2009})},\ \Eprint
  {https://arxiv.org/abs/0911.4232} {0911.4232 [gr-qc]} \BibitemShut {NoStop}%
\bibitem [{\citenamefont {Blanchet}\ \emph {et~al.}(2025)\citenamefont
  {Blanchet}, \citenamefont {Faye},\ and\ \citenamefont {Trestini}}]{BFT24}%
  \BibitemOpen
  \bibfield  {author} {\bibinfo {author} {\bibfnamefont {L.}~\bibnamefont
  {Blanchet}}, \bibinfo {author} {\bibfnamefont {G.}~\bibnamefont {Faye}},\
  and\ \bibinfo {author} {\bibfnamefont {D.}~\bibnamefont {Trestini}},\
  }\bibfield  {title} {\bibinfo {title} {{Gravitational radiation reaction for
  compact binary systems at the fourth-and-a-half post-Newtonian order}},\
  }\href {https://doi.org/10.1088/1361-6382/adac9d} {\bibfield  {journal}
  {\bibinfo  {journal} {Class. Quant. Grav.}\ }\textbf {\bibinfo {volume}
  {42}},\ \bibinfo {pages} {065015} (\bibinfo {year} {2025})},\ \Eprint
  {https://arxiv.org/abs/2407.18295} {arXiv:2407.18295 [gr-qc]} \BibitemShut
  {NoStop}%
\bibitem [{\citenamefont {Gopakumar}\ \emph {et~al.}(1997)\citenamefont
  {Gopakumar}, \citenamefont {Iyer},\ and\ \citenamefont {Iyer}}]{GII97}%
  \BibitemOpen
  \bibfield  {author} {\bibinfo {author} {\bibfnamefont {A.}~\bibnamefont
  {Gopakumar}}, \bibinfo {author} {\bibfnamefont {B.~R.}\ \bibnamefont
  {Iyer}},\ and\ \bibinfo {author} {\bibfnamefont {S.}~\bibnamefont {Iyer}},\
  }\bibfield  {title} {\bibinfo {title} {{Second post-Newtonian gravitational
  radiation reaction for two-body systems: Nonspinning bodies}},\ }\href
  {https://doi.org/10.1103/PhysRevD.57.6562} {\bibfield  {journal} {\bibinfo
  {journal} {Phys. Rev. D}\ }\textbf {\bibinfo {volume} {55}},\ \bibinfo
  {pages} {6030} (\bibinfo {year} {1997})},\ \bibinfo {note} {[Erratum:
  \href{https://doi.org/10.1103/PhysRevD.57.6562}{Phys. Rev. D 57, 6562
  (1998)}]},\ \Eprint {https://arxiv.org/abs/gr-qc/9703075}
  {arXiv:gr-qc/9703075} \BibitemShut {NoStop}%
\bibitem [{\citenamefont {Leibovich}\ \emph {et~al.}(2023)\citenamefont
  {Leibovich}, \citenamefont {Pardo},\ and\ \citenamefont
  {Yang}}]{leibovich:2023xpg}%
  \BibitemOpen
  \bibfield  {author} {\bibinfo {author} {\bibfnamefont {A.~K.}\ \bibnamefont
  {Leibovich}}, \bibinfo {author} {\bibfnamefont {B.~A.}\ \bibnamefont
  {Pardo}},\ and\ \bibinfo {author} {\bibfnamefont {Z.}~\bibnamefont {Yang}},\
  }\bibfield  {title} {\bibinfo {title} {{Radiation reaction for nonspinning
  bodies at 4.5PN in the effective field theory approach}},\ }\href
  {https://doi.org/10.1103/PhysRevD.108.024017} {\bibfield  {journal} {\bibinfo
   {journal} {Phys. Rev. D}\ }\textbf {\bibinfo {volume} {108}},\ \bibinfo
  {pages} {024017} (\bibinfo {year} {2023})},\ \Eprint
  {https://arxiv.org/abs/2302.11016} {arXiv:2302.11016 [gr-qc]} \BibitemShut
  {NoStop}%
\bibitem [{\citenamefont {Blanchet}\ \emph
  {et~al.}(2023{\natexlab{a}})\citenamefont {Blanchet}, \citenamefont {Faye},
  \citenamefont {Henry}, \citenamefont {Larrouturou},\ and\ \citenamefont
  {Trestini}}]{BFHLTa}%
  \BibitemOpen
  \bibfield  {author} {\bibinfo {author} {\bibfnamefont {L.}~\bibnamefont
  {Blanchet}}, \bibinfo {author} {\bibfnamefont {G.}~\bibnamefont {Faye}},
  \bibinfo {author} {\bibfnamefont {Q.}~\bibnamefont {Henry}}, \bibinfo
  {author} {\bibfnamefont {F.}~\bibnamefont {Larrouturou}},\ and\ \bibinfo
  {author} {\bibfnamefont {D.}~\bibnamefont {Trestini}},\ }\bibfield  {title}
  {\bibinfo {title} {{Gravitational-Wave Phasing of Quasicircular Compact
  Binary Systems to the Fourth-and-a-Half Post-Newtonian Order}},\ }\href
  {https://doi.org/10.1103/PhysRevLett.131.121402} {\bibfield  {journal}
  {\bibinfo  {journal} {Phys. Rev. Lett.}\ }\textbf {\bibinfo {volume} {131}},\
  \bibinfo {pages} {121402} (\bibinfo {year} {2023}{\natexlab{a}})},\ \Eprint
  {https://arxiv.org/abs/2304.11185} {arXiv:2304.11185 [gr-qc]} \BibitemShut
  {NoStop}%
\bibitem [{\citenamefont {Blanchet}\ \emph
  {et~al.}(2023{\natexlab{b}})\citenamefont {Blanchet}, \citenamefont {Faye},
  \citenamefont {Henry}, \citenamefont {Larrouturou},\ and\ \citenamefont
  {Trestini}}]{BFHLTb}%
  \BibitemOpen
  \bibfield  {author} {\bibinfo {author} {\bibfnamefont {L.}~\bibnamefont
  {Blanchet}}, \bibinfo {author} {\bibfnamefont {G.}~\bibnamefont {Faye}},
  \bibinfo {author} {\bibfnamefont {Q.}~\bibnamefont {Henry}}, \bibinfo
  {author} {\bibfnamefont {F.}~\bibnamefont {Larrouturou}},\ and\ \bibinfo
  {author} {\bibfnamefont {D.}~\bibnamefont {Trestini}},\ }\bibfield  {title}
  {\bibinfo {title} {{Gravitational-wave flux and quadrupole modes from
  quasicircular nonspinning compact binaries to the fourth post-Newtonian
  order}},\ }\href {https://doi.org/10.1103/PhysRevD.108.064041} {\bibfield
  {journal} {\bibinfo  {journal} {Phys. Rev. D}\ }\textbf {\bibinfo {volume}
  {108}},\ \bibinfo {pages} {064041} (\bibinfo {year} {2023}{\natexlab{b}})},\
  \Eprint {https://arxiv.org/abs/2304.11186} {arXiv:2304.11186 [gr-qc]}
  \BibitemShut {NoStop}%
\bibitem [{\citenamefont {Amalberti}\ \emph {et~al.}(2024)\citenamefont
  {Amalberti}, \citenamefont {Yang},\ and\ \citenamefont
  {Porto}}]{Amalberti:2024jaa}%
  \BibitemOpen
  \bibfield  {author} {\bibinfo {author} {\bibfnamefont {L.}~\bibnamefont
  {Amalberti}}, \bibinfo {author} {\bibfnamefont {Z.}~\bibnamefont {Yang}},\
  and\ \bibinfo {author} {\bibfnamefont {R.~A.}\ \bibnamefont {Porto}},\
  }\bibfield  {title} {\bibinfo {title} {{Gravitational radiation from
  inspiralling compact binaries to N3LO in the effective field theory
  approach}},\ }\href {https://doi.org/10.1103/PhysRevD.110.044046} {\bibfield
  {journal} {\bibinfo  {journal} {Phys. Rev. D}\ }\textbf {\bibinfo {volume}
  {110}},\ \bibinfo {pages} {044046} (\bibinfo {year} {2024})},\ \Eprint
  {https://arxiv.org/abs/2406.03457} {arXiv:2406.03457 [gr-qc]} \BibitemShut
  {NoStop}%
\bibitem [{\citenamefont {Will}\ and\ \citenamefont {Wiseman}(1996)}]{WW96}%
  \BibitemOpen
  \bibfield  {author} {\bibinfo {author} {\bibfnamefont {C.}~\bibnamefont
  {Will}}\ and\ \bibinfo {author} {\bibfnamefont {A.}~\bibnamefont {Wiseman}},\
  }\bibfield  {title} {\bibinfo {title} {{Gravitational radiation from compact
  binary systems: Gravitational waveforms and energy loss to second
  post-Newtonian order}},\ }\href {https://doi.org/10.1103/PhysRevD.54.4813}
  {\bibfield  {journal} {\bibinfo  {journal} {Phys. Rev. D}\ }\textbf {\bibinfo
  {volume} {54}},\ \bibinfo {pages} {4813} (\bibinfo {year} {1996})},\ \Eprint
  {https://arxiv.org/abs/gr-qc/9608012} {arXiv:gr-qc/9608012 [gr-qc]}
  \BibitemShut {NoStop}%
\bibitem [{\citenamefont {Le~Tiec}\ \emph {et~al.}(2012)\citenamefont
  {Le~Tiec}, \citenamefont {Blanchet},\ and\ \citenamefont {Whiting}}]{LBW12}%
  \BibitemOpen
  \bibfield  {author} {\bibinfo {author} {\bibfnamefont {A.}~\bibnamefont
  {Le~Tiec}}, \bibinfo {author} {\bibfnamefont {L.}~\bibnamefont {Blanchet}},\
  and\ \bibinfo {author} {\bibfnamefont {B.}~\bibnamefont {Whiting}},\
  }\bibfield  {title} {\bibinfo {title} {{The first law of binary black hole
  mechanics in general relativity and post-Newtonian theory}},\ }\href@noop {}
  {\bibfield  {journal} {\bibinfo  {journal} {Phys. Rev. D}\ }\textbf {\bibinfo
  {volume} {85}},\ \bibinfo {pages} {064039} (\bibinfo {year} {2012})},\
  \Eprint {https://arxiv.org/abs/1111.5378} {arXiv:1111.5378 [gr-qc]}
  \BibitemShut {NoStop}%
\bibitem [{\citenamefont {Arnowitt}\ \emph {et~al.}(1962)\citenamefont
  {Arnowitt}, \citenamefont {Deser},\ and\ \citenamefont {Misner}}]{ADM}%
  \BibitemOpen
  \bibfield  {author} {\bibinfo {author} {\bibfnamefont {R.}~\bibnamefont
  {Arnowitt}}, \bibinfo {author} {\bibfnamefont {S.}~\bibnamefont {Deser}},\
  and\ \bibinfo {author} {\bibfnamefont {C.}~\bibnamefont {Misner}},\
  }\bibfield  {title} {\bibinfo {title} {{The dynamics of General
  Relativity}},\ }in\ \href@noop {} {\emph {\bibinfo {booktitle} {An
  introduction to current research}}},\ \bibinfo {editor} {edited by\ \bibinfo
  {editor} {\bibfnamefont {L.}~\bibnamefont {Witten}}}\ (\bibinfo  {publisher}
  {New York: Wiley},\ \bibinfo {year} {1962})\BibitemShut {NoStop}%
\bibitem [{\citenamefont {Schott}(1915)}]{Schott}%
  \BibitemOpen
  \bibfield  {author} {\bibinfo {author} {\bibfnamefont {G.}~\bibnamefont
  {Schott}},\ }\bibfield  {title} {\bibinfo {title} {{On the motion of the
  Lorentz electron}},\ }\href
  {https://doi.org/https://doi.org/10.1080/14786440108635280} {\bibfield
  {journal} {\bibinfo  {journal} {Phil. Mag.}\ }\bibinfo {series} {6},\ \textbf
  {\bibinfo {volume} {29}},\ \bibinfo {pages} {49} (\bibinfo {year}
  {1915})}\BibitemShut {NoStop}%
\bibitem [{\citenamefont {Grant}\ \emph {et~al.}(2025)\citenamefont {Grant},
  \citenamefont {{Le Tiec}},\ and\ \citenamefont {Pound}}]{GLTP25}%
  \BibitemOpen
  \bibfield  {author} {\bibinfo {author} {\bibfnamefont {A.~M.}\ \bibnamefont
  {Grant}}, \bibinfo {author} {\bibfnamefont {A.}~\bibnamefont {{Le Tiec}}},\
  and\ \bibinfo {author} {\bibfnamefont {A.}~\bibnamefont {Pound}},\ }\bibfield
   {title} {\bibinfo {title} {The first law of binary black hole mechanics with
  dissipation},\ }\href@noop {} {\bibfield  {journal} {\bibinfo  {journal}
  {\textit{in prep.}}\ } (\bibinfo {year} {2025})}\BibitemShut {NoStop}%
\bibitem [{\citenamefont {Wald}\ and\ \citenamefont
  {Zoupas}(2000)}]{Wald:1999wa}%
  \BibitemOpen
  \bibfield  {author} {\bibinfo {author} {\bibfnamefont {R.~M.}\ \bibnamefont
  {Wald}}\ and\ \bibinfo {author} {\bibfnamefont {A.}~\bibnamefont {Zoupas}},\
  }\bibfield  {title} {\bibinfo {title} {{A General definition of 'conserved
  quantities' in general relativity and other theories of gravity}},\ }\href
  {https://doi.org/10.1103/PhysRevD.61.084027} {\bibfield  {journal} {\bibinfo
  {journal} {Phys. Rev. D}\ }\textbf {\bibinfo {volume} {61}},\ \bibinfo
  {pages} {084027} (\bibinfo {year} {2000})},\ \Eprint
  {https://arxiv.org/abs/gr-qc/9911095} {arXiv:gr-qc/9911095} \BibitemShut
  {NoStop}%
\bibitem [{\citenamefont {Grant}\ \emph {et~al.}(2022)\citenamefont {Grant},
  \citenamefont {Prabhu},\ and\ \citenamefont {Shehzad}}]{Grant:2021sxk}%
  \BibitemOpen
  \bibfield  {author} {\bibinfo {author} {\bibfnamefont {A.~M.}\ \bibnamefont
  {Grant}}, \bibinfo {author} {\bibfnamefont {K.}~\bibnamefont {Prabhu}},\ and\
  \bibinfo {author} {\bibfnamefont {I.}~\bibnamefont {Shehzad}},\ }\bibfield
  {title} {\bibinfo {title} {{The Wald\textendash{}Zoupas prescription for
  asymptotic charges at null infinity in general relativity}},\ }\href
  {https://doi.org/10.1088/1361-6382/ac571a} {\bibfield  {journal} {\bibinfo
  {journal} {Class. Quant. Grav.}\ }\textbf {\bibinfo {volume} {39}},\ \bibinfo
  {pages} {085002} (\bibinfo {year} {2022})},\ \Eprint
  {https://arxiv.org/abs/2105.05919} {arXiv:2105.05919 [gr-qc]} \BibitemShut
  {NoStop}%
\bibitem [{\citenamefont {Breuer}\ and\ \citenamefont
  {Rudolph}(1981)}]{breuer_radiation_1981}%
  \BibitemOpen
  \bibfield  {author} {\bibinfo {author} {\bibfnamefont {R.~A.}\ \bibnamefont
  {Breuer}}\ and\ \bibinfo {author} {\bibfnamefont {E.}~\bibnamefont
  {Rudolph}},\ }\bibfield  {title} {\bibinfo {title} {Radiation reaction and
  energy loss in the post-newtonian approximation of general relativity},\
  }\href {https://doi.org/10.1007/BF00758216} {\bibfield  {journal} {\bibinfo
  {journal} {General Relativity and Gravitation}\ }\textbf {\bibinfo {volume}
  {13}},\ \bibinfo {pages} {777} (\bibinfo {year} {1981})}\BibitemShut
  {NoStop}%
\bibitem [{\citenamefont {Marchand}\ \emph {et~al.}(2020)\citenamefont
  {Marchand}, \citenamefont {Henry}, \citenamefont {Larrouturou}, \citenamefont
  {Marsat}, \citenamefont {Faye},\ and\ \citenamefont {Blanchet}}]{MHLMFB20}%
  \BibitemOpen
  \bibfield  {author} {\bibinfo {author} {\bibfnamefont {T.}~\bibnamefont
  {Marchand}}, \bibinfo {author} {\bibfnamefont {Q.}~\bibnamefont {Henry}},
  \bibinfo {author} {\bibfnamefont {F.}~\bibnamefont {Larrouturou}}, \bibinfo
  {author} {\bibfnamefont {S.}~\bibnamefont {Marsat}}, \bibinfo {author}
  {\bibfnamefont {G.}~\bibnamefont {Faye}},\ and\ \bibinfo {author}
  {\bibfnamefont {L.}~\bibnamefont {Blanchet}},\ }\bibfield  {title} {\bibinfo
  {title} {{The mass quadrupole moment of compact binary systems at the fourth
  post-Newtonian order}},\ }\href {https://doi.org/10.1088/1361-6382/ab9ce1}
  {\bibfield  {journal} {\bibinfo  {journal} {Class. Quant. Grav.}\ }\textbf
  {\bibinfo {volume} {37}},\ \bibinfo {pages} {215006} (\bibinfo {year}
  {2020})},\ \Eprint {https://arxiv.org/abs/2003.13672} {arXiv:2003.13672
  [gr-qc]} \BibitemShut {NoStop}%
\bibitem [{\citenamefont {Larrouturou}\ \emph
  {et~al.}(2022{\natexlab{a}})\citenamefont {Larrouturou}, \citenamefont
  {Henry}, \citenamefont {Blanchet},\ and\ \citenamefont {Faye}}]{MQ4PN_IR}%
  \BibitemOpen
  \bibfield  {author} {\bibinfo {author} {\bibfnamefont {F.}~\bibnamefont
  {Larrouturou}}, \bibinfo {author} {\bibfnamefont {Q.}~\bibnamefont {Henry}},
  \bibinfo {author} {\bibfnamefont {L.}~\bibnamefont {Blanchet}},\ and\
  \bibinfo {author} {\bibfnamefont {G.}~\bibnamefont {Faye}},\ }\bibfield
  {title} {\bibinfo {title} {{The quadrupole moment of compact binaries to the
  fourth post-Newtonian order: I. Non-locality in time and infra-red
  divergencies}},\ }\href {https://doi.org/10.1088/1361-6382/ac5762} {\bibfield
   {journal} {\bibinfo  {journal} {Class. Quant. Grav.}\ }\textbf {\bibinfo
  {volume} {39}},\ \bibinfo {pages} {115007} (\bibinfo {year}
  {2022}{\natexlab{a}})},\ \Eprint {https://arxiv.org/abs/2110.02240}
  {arXiv:2110.02240 [gr-qc]} \BibitemShut {NoStop}%
\bibitem [{\citenamefont {Larrouturou}\ \emph
  {et~al.}(2022{\natexlab{b}})\citenamefont {Larrouturou}, \citenamefont
  {Blanchet}, \citenamefont {Henry},\ and\ \citenamefont
  {Faye}}]{MQ4PN_renorm}%
  \BibitemOpen
  \bibfield  {author} {\bibinfo {author} {\bibfnamefont {F.}~\bibnamefont
  {Larrouturou}}, \bibinfo {author} {\bibfnamefont {L.}~\bibnamefont
  {Blanchet}}, \bibinfo {author} {\bibfnamefont {Q.}~\bibnamefont {Henry}},\
  and\ \bibinfo {author} {\bibfnamefont {G.}~\bibnamefont {Faye}},\ }\bibfield
  {title} {\bibinfo {title} {{The quadrupole moment of compact binaries to the
  fourth post-Newtonian order: II. Dimensional regularization and
  renormalization}},\ }\href {https://doi.org/10.1088/1361-6382/ac5ba0}
  {\bibfield  {journal} {\bibinfo  {journal} {Class. Quant. Grav.}\ }\textbf
  {\bibinfo {volume} {39}},\ \bibinfo {pages} {115008} (\bibinfo {year}
  {2022}{\natexlab{b}})},\ \Eprint {https://arxiv.org/abs/2110.02243}
  {arXiv:2110.02243 [gr-qc]} \BibitemShut {NoStop}%
\bibitem [{\citenamefont {Blanchet}\ \emph {et~al.}(2022)\citenamefont
  {Blanchet}, \citenamefont {Faye},\ and\ \citenamefont
  {Larrouturou}}]{MQ4PN_jauge}%
  \BibitemOpen
  \bibfield  {author} {\bibinfo {author} {\bibfnamefont {L.}~\bibnamefont
  {Blanchet}}, \bibinfo {author} {\bibfnamefont {G.}~\bibnamefont {Faye}},\
  and\ \bibinfo {author} {\bibfnamefont {F.}~\bibnamefont {Larrouturou}},\
  }\bibfield  {title} {\bibinfo {title} {{The quadrupole moment of compact
  binaries to the fourth post-Newtonian order: from source to canonical
  moment}},\ }\href {https://doi.org/10.1088/1361-6382/ac840c} {\bibfield
  {journal} {\bibinfo  {journal} {Class. Quant. Grav.}\ }\textbf {\bibinfo
  {volume} {39}},\ \bibinfo {pages} {195003} (\bibinfo {year} {2022})},\
  \Eprint {https://arxiv.org/abs/2204.11293} {arXiv:2204.11293 [gr-qc]}
  \BibitemShut {NoStop}%
\bibitem [{\citenamefont {Henry}\ \emph {et~al.}(2021)\citenamefont {Henry},
  \citenamefont {Faye},\ and\ \citenamefont {Blanchet}}]{Henry:2021cek}%
  \BibitemOpen
  \bibfield  {author} {\bibinfo {author} {\bibfnamefont {Q.}~\bibnamefont
  {Henry}}, \bibinfo {author} {\bibfnamefont {G.}~\bibnamefont {Faye}},\ and\
  \bibinfo {author} {\bibfnamefont {L.}~\bibnamefont {Blanchet}},\ }\bibfield
  {title} {\bibinfo {title} {{The current-type quadrupole moment and
  gravitational-wave mode \mbox{(\ensuremath{\ell}, m) = (2, 1)} of compact
  binary systems at the third post-{N}ewtonian order}},\ }\href
  {https://doi.org/10.1088/1361-6382/ac1850} {\bibfield  {journal} {\bibinfo
  {journal} {Class. Quant. Grav.}\ }\textbf {\bibinfo {volume} {38}},\ \bibinfo
  {pages} {185004} (\bibinfo {year} {2021})},\ \Eprint
  {https://arxiv.org/abs/2105.10876} {arXiv:2105.10876 [gr-qc]} \BibitemShut
  {NoStop}%
\bibitem [{\citenamefont {Trestini}\ \emph {et~al.}(2023)\citenamefont
  {Trestini}, \citenamefont {Larrouturou},\ and\ \citenamefont
  {Blanchet}}]{TLB22}%
  \BibitemOpen
  \bibfield  {author} {\bibinfo {author} {\bibfnamefont {D.}~\bibnamefont
  {Trestini}}, \bibinfo {author} {\bibfnamefont {F.}~\bibnamefont
  {Larrouturou}},\ and\ \bibinfo {author} {\bibfnamefont {L.}~\bibnamefont
  {Blanchet}},\ }\bibfield  {title} {\bibinfo {title} {{The quadrupole moment
  of compact binaries to the fourth post-Newtonian order: relating the harmonic
  and radiative metrics}},\ }\href {https://doi.org/10.1088/1361-6382/acb5de}
  {\bibfield  {journal} {\bibinfo  {journal} {Class. Quant. Grav.}\ }\textbf
  {\bibinfo {volume} {40}},\ \bibinfo {pages} {055006} (\bibinfo {year}
  {2023})},\ \Eprint {https://arxiv.org/abs/2209.02719} {arXiv:2209.02719
  [gr-qc]} \BibitemShut {NoStop}%
\bibitem [{\citenamefont {Trestini}\ and\ \citenamefont
  {Blanchet}(2023)}]{TB23_ToM}%
  \BibitemOpen
  \bibfield  {author} {\bibinfo {author} {\bibfnamefont {D.}~\bibnamefont
  {Trestini}}\ and\ \bibinfo {author} {\bibfnamefont {L.}~\bibnamefont
  {Blanchet}},\ }\bibfield  {title} {\bibinfo {title} {{Gravitational-wave
  tails of memory}},\ }\href {https://doi.org/10.1103/PhysRevD.107.104048}
  {\bibfield  {journal} {\bibinfo  {journal} {Phys. Rev. D}\ }\textbf {\bibinfo
  {volume} {107}},\ \bibinfo {pages} {104048} (\bibinfo {year} {2023})},\
  \Eprint {https://arxiv.org/abs/2301.09395} {arXiv:2301.09395 [gr-qc]}
  \BibitemShut {NoStop}%
\bibitem [{\citenamefont {Blanchet}(1987)}]{B87}%
  \BibitemOpen
  \bibfield  {author} {\bibinfo {author} {\bibfnamefont {L.}~\bibnamefont
  {Blanchet}},\ }\bibfield  {title} {\bibinfo {title} {{Radiative gravitational
  fields in general relativity. II.~Asymptotic behaviour at future null
  infinity}},\ }\href {https://doi.org/10.1098/rspa.1987.0022} {\bibfield
  {journal} {\bibinfo  {journal} {Proc. Roy. Soc. Lond. A}\ }\textbf {\bibinfo
  {volume} {409}},\ \bibinfo {pages} {383} (\bibinfo {year}
  {1987})}\BibitemShut {NoStop}%
\bibitem [{\citenamefont {Detweiler}(2005)}]{Detweiler:2005kq}%
  \BibitemOpen
  \bibfield  {author} {\bibinfo {author} {\bibfnamefont {S.~L.}\ \bibnamefont
  {Detweiler}},\ }\bibfield  {title} {\bibinfo {title} {{Perspective on
  gravitational self-force analyses}},\ }\href
  {https://doi.org/10.1088/0264-9381/22/15/006} {\bibfield  {journal} {\bibinfo
   {journal} {Class. Quant. Grav.}\ }\textbf {\bibinfo {volume} {22}},\
  \bibinfo {pages} {S681} (\bibinfo {year} {2005})},\ \Eprint
  {https://arxiv.org/abs/gr-qc/0501004} {arXiv:gr-qc/0501004} \BibitemShut
  {NoStop}%
\bibitem [{\citenamefont {Blanchet}\ and\ \citenamefont {Damour}(1988)}]{BD88}%
  \BibitemOpen
  \bibfield  {author} {\bibinfo {author} {\bibfnamefont {L.}~\bibnamefont
  {Blanchet}}\ and\ \bibinfo {author} {\bibfnamefont {T.}~\bibnamefont
  {Damour}},\ }\bibfield  {title} {\bibinfo {title} {{Tail Transported Temporal
  Correlations in the Dynamics of a Gravitating System}},\ }\href
  {https://doi.org/10.1103/PhysRevD.37.1410} {\bibfield  {journal} {\bibinfo
  {journal} {Phys. Rev. D}\ }\textbf {\bibinfo {volume} {37}},\ \bibinfo
  {pages} {1410} (\bibinfo {year} {1988})}\BibitemShut {NoStop}%
\bibitem [{\citenamefont {Blanchet}\ and\ \citenamefont {Damour}(1992)}]{BD92}%
  \BibitemOpen
  \bibfield  {author} {\bibinfo {author} {\bibfnamefont {L.}~\bibnamefont
  {Blanchet}}\ and\ \bibinfo {author} {\bibfnamefont {T.}~\bibnamefont
  {Damour}},\ }\bibfield  {title} {\bibinfo {title} {{Hereditary effects in
  gravitational radiation}},\ }\href {https://doi.org/10.1103/PhysRevD.46.4304}
  {\bibfield  {journal} {\bibinfo  {journal} {Phys. Rev. D}\ }\textbf {\bibinfo
  {volume} {46}},\ \bibinfo {pages} {4304} (\bibinfo {year}
  {1992})}\BibitemShut {NoStop}%
\bibitem [{\citenamefont {Blanchet}(1993)}]{B93}%
  \BibitemOpen
  \bibfield  {author} {\bibinfo {author} {\bibfnamefont {L.}~\bibnamefont
  {Blanchet}},\ }\bibfield  {title} {\bibinfo {title} {{Time asymmetric
  structure of gravitational radiation}},\ }\href
  {https://doi.org/10.1103/PhysRevD.47.4392} {\bibfield  {journal} {\bibinfo
  {journal} {Phys. Rev. D}\ }\textbf {\bibinfo {volume} {47}},\ \bibinfo
  {pages} {4392} (\bibinfo {year} {1993})}\BibitemShut {NoStop}%
\bibitem [{\citenamefont {Blanchet}(1997)}]{B97}%
  \BibitemOpen
  \bibfield  {author} {\bibinfo {author} {\bibfnamefont {L.}~\bibnamefont
  {Blanchet}},\ }\bibfield  {title} {\bibinfo {title} {{Gravitational radiation
  reaction and balance equations to post-Newtonian order}},\ }\href
  {https://doi.org/10.1103/PhysRevD.55.714} {\bibfield  {journal} {\bibinfo
  {journal} {Phys. Rev. D}\ }\textbf {\bibinfo {volume} {55}},\ \bibinfo
  {pages} {714} (\bibinfo {year} {1997})},\ \Eprint
  {https://arxiv.org/abs/gr-qc/9609049} {arXiv:gr-qc/9609049} \BibitemShut
  {NoStop}%
\bibitem [{\citenamefont {Blanchet}(2014)}]{Blanchet:2013haa}%
  \BibitemOpen
  \bibfield  {author} {\bibinfo {author} {\bibfnamefont {L.}~\bibnamefont
  {Blanchet}},\ }\bibfield  {title} {\bibinfo {title} {{Post-Newtonian Theory
  for Gravitational Waves}},\ }\href {https://doi.org/10.12942/lrr-2014-2}
  {\bibfield  {journal} {\bibinfo  {journal} {Living Rev. Rel.}\ }\textbf
  {\bibinfo {volume} {17}},\ \bibinfo {pages} {2} (\bibinfo {year} {2014})},\
  \Eprint {https://arxiv.org/abs/1310.1528} {arXiv:1310.1528 [gr-qc]}
  \BibitemShut {NoStop}%
\bibitem [{\citenamefont {Galley}\ \emph {et~al.}(2016)\citenamefont {Galley},
  \citenamefont {Leibovich}, \citenamefont {Porto},\ and\ \citenamefont
  {Ross}}]{GLPR16}%
  \BibitemOpen
  \bibfield  {author} {\bibinfo {author} {\bibfnamefont {C.~R.}\ \bibnamefont
  {Galley}}, \bibinfo {author} {\bibfnamefont {A.~K.}\ \bibnamefont
  {Leibovich}}, \bibinfo {author} {\bibfnamefont {R.~A.}\ \bibnamefont
  {Porto}},\ and\ \bibinfo {author} {\bibfnamefont {A.}~\bibnamefont {Ross}},\
  }\bibfield  {title} {\bibinfo {title} {{Tail effect in gravitational
  radiation reaction: Time nonlocality and renormalization group evolution}},\
  }\href {https://doi.org/10.1103/PhysRevD.93.124010} {\bibfield  {journal}
  {\bibinfo  {journal} {Phys. Rev. D}\ }\textbf {\bibinfo {volume} {93}},\
  \bibinfo {pages} {124010} (\bibinfo {year} {2016})},\ \Eprint
  {https://arxiv.org/abs/1511.07379} {arXiv:1511.07379 [gr-qc]} \BibitemShut
  {NoStop}%
\bibitem [{\citenamefont {Blanchet}\ and\ \citenamefont
  {Le~Tiec}(2017)}]{Blanchet:2017rcn}%
  \BibitemOpen
  \bibfield  {author} {\bibinfo {author} {\bibfnamefont {L.}~\bibnamefont
  {Blanchet}}\ and\ \bibinfo {author} {\bibfnamefont {A.}~\bibnamefont
  {Le~Tiec}},\ }\bibfield  {title} {\bibinfo {title} {{First Law of Compact
  Binary Mechanics with Gravitational-Wave Tails}},\ }\href
  {https://doi.org/10.1088/1361-6382/aa79d7} {\bibfield  {journal} {\bibinfo
  {journal} {Class. Quant. Grav.}\ }\textbf {\bibinfo {volume} {34}},\ \bibinfo
  {pages} {164001} (\bibinfo {year} {2017})},\ \Eprint
  {https://arxiv.org/abs/1702.06839} {arXiv:1702.06839 [gr-qc]} \BibitemShut
  {NoStop}%
\bibitem [{\citenamefont {Thorne}(1980)}]{Thorne:1980ru}%
  \BibitemOpen
  \bibfield  {author} {\bibinfo {author} {\bibfnamefont {K.~S.}\ \bibnamefont
  {Thorne}},\ }\bibfield  {title} {\bibinfo {title} {{Multipole Expansions of
  Gravitational Radiation}},\ }\href
  {https://doi.org/10.1103/RevModPhys.52.299} {\bibfield  {journal} {\bibinfo
  {journal} {Rev. Mod. Phys.}\ }\textbf {\bibinfo {volume} {52}},\ \bibinfo
  {pages} {299} (\bibinfo {year} {1980})}\BibitemShut {NoStop}%
\bibitem [{\citenamefont {Blanchet}\ and\ \citenamefont {Faye}(2019)}]{BF18}%
  \BibitemOpen
  \bibfield  {author} {\bibinfo {author} {\bibfnamefont {L.}~\bibnamefont
  {Blanchet}}\ and\ \bibinfo {author} {\bibfnamefont {G.}~\bibnamefont
  {Faye}},\ }\bibfield  {title} {\bibinfo {title} {{Flux-balance equations for
  linear momentum and center-of-mass position of self-gravitating
  post-Newtonian systems}},\ }\href {https://doi.org/10.1088/1361-6382/ab0d4f}
  {\bibfield  {journal} {\bibinfo  {journal} {Class. Quant. Grav.}\ }\textbf
  {\bibinfo {volume} {36}},\ \bibinfo {pages} {085003} (\bibinfo {year}
  {2019})},\ \Eprint {https://arxiv.org/abs/1811.08966} {arXiv:1811.08966
  [gr-qc]} \BibitemShut {NoStop}%
\bibitem [{\citenamefont {Comp\`ere}\ \emph {et~al.}(2020)\citenamefont
  {Comp\`ere}, \citenamefont {Oliveri},\ and\ \citenamefont {Seraj}}]{COS19}%
  \BibitemOpen
  \bibfield  {author} {\bibinfo {author} {\bibfnamefont {G.}~\bibnamefont
  {Comp\`ere}}, \bibinfo {author} {\bibfnamefont {R.}~\bibnamefont {Oliveri}},\
  and\ \bibinfo {author} {\bibfnamefont {A.}~\bibnamefont {Seraj}},\ }\bibfield
   {title} {\bibinfo {title} {{The Poincar\'e and BMS flux-balance laws with
  application to binary systems}},\ }\href
  {https://doi.org/10.1007/JHEP10(2020)116} {\bibfield  {journal} {\bibinfo
  {journal} {JHEP}\ }\textbf {\bibinfo {volume} {10}},\ \bibinfo {pages}
  {116}},\ \Eprint {https://arxiv.org/abs/1912.03164} {arXiv:1912.03164
  [gr-qc]} \BibitemShut {NoStop}%
\bibitem [{\citenamefont {Blanchet}\ and\ \citenamefont
  {Sch{\"a}fer}(1993)}]{BS93}%
  \BibitemOpen
  \bibfield  {author} {\bibinfo {author} {\bibfnamefont {L.}~\bibnamefont
  {Blanchet}}\ and\ \bibinfo {author} {\bibfnamefont {G.}~\bibnamefont
  {Sch{\"a}fer}},\ }\bibfield  {title} {\bibinfo {title} {{Gravitational wave
  tails and binary star systems}},\ }\href
  {https://doi.org/10.1088/0264-9381/10/12/026} {\bibfield  {journal} {\bibinfo
   {journal} {Class. Quant. Grav.}\ }\textbf {\bibinfo {volume} {10}},\
  \bibinfo {pages} {2699} (\bibinfo {year} {1993})}\BibitemShut {NoStop}%
\bibitem [{\citenamefont {Arun}\ \emph {et~al.}(2004)\citenamefont {Arun},
  \citenamefont {Blanchet}, \citenamefont {Iyer},\ and\ \citenamefont
  {Qusailah}}]{ABIQ04}%
  \BibitemOpen
  \bibfield  {author} {\bibinfo {author} {\bibfnamefont {K.~G.}\ \bibnamefont
  {Arun}}, \bibinfo {author} {\bibfnamefont {L.}~\bibnamefont {Blanchet}},
  \bibinfo {author} {\bibfnamefont {B.~R.}\ \bibnamefont {Iyer}},\ and\
  \bibinfo {author} {\bibfnamefont {M.~S.~S.}\ \bibnamefont {Qusailah}},\
  }\bibfield  {title} {\bibinfo {title} {{The 2.5PN gravitational wave
  polarisations from inspiralling compact binaries in circular orbits}},\
  }\href {https://doi.org/10.1088/0264-9381/21/15/010} {\bibfield  {journal}
  {\bibinfo  {journal} {Class. Quant. Grav.}\ }\textbf {\bibinfo {volume}
  {21}},\ \bibinfo {pages} {3771} (\bibinfo {year} {2004})},\ \bibinfo {note}
  {{[Corrigendum: \href{https://doi.org/10.1088/0264-9381/22/14/C01}{Class.
  Quant. Grav. $\bm{22}$, 3115 (2005)}]}},\ \Eprint
  {https://arxiv.org/abs/gr-qc/0404085} {arXiv:gr-qc/0404085} \BibitemShut
  {NoStop}%
\bibitem [{\citenamefont {Trestini}(2023)}]{Trestini:2023khz}%
  \BibitemOpen
  \bibfield  {author} {\bibinfo {author} {\bibfnamefont {D.}~\bibnamefont
  {Trestini}},\ }\emph {\bibinfo {title} {{Gravitational radiation of compact
  binary systems in general relativity and in scalar-tensor theories}}},\ \href
  {https://theses.hal.science/tel-04224762} {Ph.D. thesis},\ \bibinfo  {school}
  {Institut d'Astrophysique de Paris, France, Sorbonne Universit\'e} (\bibinfo
  {year} {2023}),\ \bibinfo {note} {tel-04224762}\BibitemShut {NoStop}%
\bibitem [{\citenamefont {Khairnar}\ \emph {et~al.}(2025)\citenamefont
  {Khairnar}, \citenamefont {Stein},\ and\ \citenamefont
  {Boyle}}]{Khairnar:2024rzs}%
  \BibitemOpen
  \bibfield  {author} {\bibinfo {author} {\bibfnamefont {A.}~\bibnamefont
  {Khairnar}}, \bibinfo {author} {\bibfnamefont {L.~C.}\ \bibnamefont
  {Stein}},\ and\ \bibinfo {author} {\bibfnamefont {M.}~\bibnamefont {Boyle}},\
  }\bibfield  {title} {\bibinfo {title} {{Approximate helical symmetry in
  compact binaries}},\ }\href {https://doi.org/10.1103/PhysRevD.111.024072}
  {\bibfield  {journal} {\bibinfo  {journal} {Phys. Rev. D}\ }\textbf {\bibinfo
  {volume} {111}},\ \bibinfo {pages} {024072} (\bibinfo {year} {2025})},\
  \Eprint {https://arxiv.org/abs/2410.16373} {arXiv:2410.16373 [gr-qc]}
  \BibitemShut {NoStop}%
\bibitem [{\citenamefont {\"Ozel}\ and\ \citenamefont
  {Freire}(2016)}]{Ozel:2016oaf}%
  \BibitemOpen
  \bibfield  {author} {\bibinfo {author} {\bibfnamefont {F.}~\bibnamefont
  {\"Ozel}}\ and\ \bibinfo {author} {\bibfnamefont {P.}~\bibnamefont
  {Freire}},\ }\bibfield  {title} {\bibinfo {title} {{Masses, Radii, and the
  Equation of State of Neutron Stars}},\ }\href
  {https://doi.org/10.1146/annurev-astro-081915-023322} {\bibfield  {journal}
  {\bibinfo  {journal} {Ann. Rev. Astron. Astrophys.}\ }\textbf {\bibinfo
  {volume} {54}},\ \bibinfo {pages} {401} (\bibinfo {year} {2016})},\ \Eprint
  {https://arxiv.org/abs/1603.02698} {arXiv:1603.02698 [astro-ph.HE]}
  \BibitemShut {NoStop}%
\bibitem [{\citenamefont {Taylor}\ and\ \citenamefont
  {Poisson}(2008)}]{Taylor:2008xy}%
  \BibitemOpen
  \bibfield  {author} {\bibinfo {author} {\bibfnamefont {S.}~\bibnamefont
  {Taylor}}\ and\ \bibinfo {author} {\bibfnamefont {E.}~\bibnamefont
  {Poisson}},\ }\bibfield  {title} {\bibinfo {title} {{Nonrotating black hole
  in a post-Newtonian tidal environment}},\ }\href
  {https://doi.org/10.1103/PhysRevD.78.084016} {\bibfield  {journal} {\bibinfo
  {journal} {Phys. Rev. D}\ }\textbf {\bibinfo {volume} {78}},\ \bibinfo
  {pages} {084016} (\bibinfo {year} {2008})},\ \Eprint
  {https://arxiv.org/abs/0806.3052} {arXiv:0806.3052 [gr-qc]} \BibitemShut
  {NoStop}%
\bibitem [{\citenamefont {Nagar}\ and\ \citenamefont
  {Akcay}(2012)}]{Nagar:2011aa}%
  \BibitemOpen
  \bibfield  {author} {\bibinfo {author} {\bibfnamefont {A.}~\bibnamefont
  {Nagar}}\ and\ \bibinfo {author} {\bibfnamefont {S.}~\bibnamefont {Akcay}},\
  }\bibfield  {title} {\bibinfo {title} {{Horizon-absorbed energy flux in
  circularized, nonspinning black-hole binaries and its effective-one-body
  representation}},\ }\href {https://doi.org/10.1103/PhysRevD.85.044025}
  {\bibfield  {journal} {\bibinfo  {journal} {Phys. Rev. D}\ }\textbf {\bibinfo
  {volume} {85}},\ \bibinfo {pages} {044025} (\bibinfo {year} {2012})},\
  \Eprint {https://arxiv.org/abs/1112.2840} {arXiv:1112.2840 [gr-qc]}
  \BibitemShut {NoStop}%
\bibitem [{\citenamefont {Saketh}\ \emph {et~al.}(2024)\citenamefont {Saketh},
  \citenamefont {Zhou}, \citenamefont {Ghosh}, \citenamefont {Steinhoff},\ and\
  \citenamefont {Chatterjee}}]{Saketh:2024juq}%
  \BibitemOpen
  \bibfield  {author} {\bibinfo {author} {\bibfnamefont {M.~V.~S.}\
  \bibnamefont {Saketh}}, \bibinfo {author} {\bibfnamefont {Z.}~\bibnamefont
  {Zhou}}, \bibinfo {author} {\bibfnamefont {S.}~\bibnamefont {Ghosh}},
  \bibinfo {author} {\bibfnamefont {J.}~\bibnamefont {Steinhoff}},\ and\
  \bibinfo {author} {\bibfnamefont {D.}~\bibnamefont {Chatterjee}},\ }\bibfield
   {title} {\bibinfo {title} {{Investigating tidal heating in neutron stars via
  gravitational Raman scattering}},\ }\href
  {https://doi.org/10.1103/PhysRevD.110.103001} {\bibfield  {journal} {\bibinfo
   {journal} {Phys. Rev. D}\ }\textbf {\bibinfo {volume} {110}},\ \bibinfo
  {pages} {103001} (\bibinfo {year} {2024})},\ \Eprint
  {https://arxiv.org/abs/2407.08327} {arXiv:2407.08327 [gr-qc]} \BibitemShut
  {NoStop}%
\bibitem [{\citenamefont {Warburton}\ \emph {et~al.}(2024)\citenamefont
  {Warburton}, \citenamefont {Wardell}, \citenamefont {Trestini}, \citenamefont
  {Henry}, \citenamefont {Pound}, \citenamefont {Blanchet}, \citenamefont
  {Durkan}, \citenamefont {Faye},\ and\ \citenamefont
  {Miller}}]{Warburton:2024xnr}%
  \BibitemOpen
  \bibfield  {author} {\bibinfo {author} {\bibfnamefont {N.}~\bibnamefont
  {Warburton}}, \bibinfo {author} {\bibfnamefont {B.}~\bibnamefont {Wardell}},
  \bibinfo {author} {\bibfnamefont {D.}~\bibnamefont {Trestini}}, \bibinfo
  {author} {\bibfnamefont {Q.}~\bibnamefont {Henry}}, \bibinfo {author}
  {\bibfnamefont {A.}~\bibnamefont {Pound}}, \bibinfo {author} {\bibfnamefont
  {L.}~\bibnamefont {Blanchet}}, \bibinfo {author} {\bibfnamefont
  {L.}~\bibnamefont {Durkan}}, \bibinfo {author} {\bibfnamefont
  {G.}~\bibnamefont {Faye}},\ and\ \bibinfo {author} {\bibfnamefont
  {J.}~\bibnamefont {Miller}},\ }\bibfield  {title} {\bibinfo {title}
  {{Comparison of 4.5PN and 2SF gravitational energy fluxes from quasicircular
  compact binaries}},\ }\href@noop {} {\  (\bibinfo {year} {2024})},\ \Eprint
  {https://arxiv.org/abs/2407.00366} {arXiv:2407.00366 [gr-qc]} \BibitemShut
  {NoStop}%
\bibitem [{\citenamefont {Nagar}\ \emph {et~al.}(2024)\citenamefont {Nagar},
  \citenamefont {Gamba}, \citenamefont {Rettegno}, \citenamefont {Fantini},\
  and\ \citenamefont {Bernuzzi}}]{Nagar:2024dzj}%
  \BibitemOpen
  \bibfield  {author} {\bibinfo {author} {\bibfnamefont {A.}~\bibnamefont
  {Nagar}}, \bibinfo {author} {\bibfnamefont {R.}~\bibnamefont {Gamba}},
  \bibinfo {author} {\bibfnamefont {P.}~\bibnamefont {Rettegno}}, \bibinfo
  {author} {\bibfnamefont {V.}~\bibnamefont {Fantini}},\ and\ \bibinfo {author}
  {\bibfnamefont {S.}~\bibnamefont {Bernuzzi}},\ }\bibfield  {title} {\bibinfo
  {title} {{Effective-one-body waveform model for noncircularized, planar,
  coalescing black hole binaries: The importance of radiation reaction}},\
  }\href {https://doi.org/10.1103/PhysRevD.110.084001} {\bibfield  {journal}
  {\bibinfo  {journal} {Phys. Rev. D}\ }\textbf {\bibinfo {volume} {110}},\
  \bibinfo {pages} {084001} (\bibinfo {year} {2024})},\ \Eprint
  {https://arxiv.org/abs/2404.05288} {arXiv:2404.05288 [gr-qc]} \BibitemShut
  {NoStop}%
\bibitem [{\citenamefont {Nagar}\ \emph {et~al.}(2025)\citenamefont {Nagar},
  \citenamefont {Chiaramello}, \citenamefont {Gamba}, \citenamefont {Albanesi},
  \citenamefont {Bernuzzi}, \citenamefont {Fantini}, \citenamefont {Panzeri},\
  and\ \citenamefont {Rettegno}}]{Nagar:2024oyk}%
  \BibitemOpen
  \bibfield  {author} {\bibinfo {author} {\bibfnamefont {A.}~\bibnamefont
  {Nagar}}, \bibinfo {author} {\bibfnamefont {D.}~\bibnamefont {Chiaramello}},
  \bibinfo {author} {\bibfnamefont {R.}~\bibnamefont {Gamba}}, \bibinfo
  {author} {\bibfnamefont {S.}~\bibnamefont {Albanesi}}, \bibinfo {author}
  {\bibfnamefont {S.}~\bibnamefont {Bernuzzi}}, \bibinfo {author}
  {\bibfnamefont {V.}~\bibnamefont {Fantini}}, \bibinfo {author} {\bibfnamefont
  {M.}~\bibnamefont {Panzeri}},\ and\ \bibinfo {author} {\bibfnamefont
  {P.}~\bibnamefont {Rettegno}},\ }\bibfield  {title} {\bibinfo {title}
  {{Effective-one-body waveform model for noncircularized, planar, coalescing
  black hole binaries. II. High accuracy by improving logarithmic terms in
  resummations}},\ }\href {https://doi.org/10.1103/PhysRevD.111.064050}
  {\bibfield  {journal} {\bibinfo  {journal} {Phys. Rev. D}\ }\textbf {\bibinfo
  {volume} {111}},\ \bibinfo {pages} {064050} (\bibinfo {year} {2025})},\
  \Eprint {https://arxiv.org/abs/2407.04762} {arXiv:2407.04762 [gr-qc]}
  \BibitemShut {NoStop}%
\bibitem [{\citenamefont {Fujita}\ and\ \citenamefont {Iyer}(2010)}]{FI10}%
  \BibitemOpen
  \bibfield  {author} {\bibinfo {author} {\bibfnamefont {R.}~\bibnamefont
  {Fujita}}\ and\ \bibinfo {author} {\bibfnamefont {B.~R.}\ \bibnamefont
  {Iyer}},\ }\bibfield  {title} {\bibinfo {title} {{Spherical harmonic modes of
  5.5 post-Newtonian gravitational wave polarisations and associated factorised
  resummed waveforms for a particle in circular orbit around a Schwarzschild
  black hole}},\ }\href {https://doi.org/10.1103/PhysRevD.82.044051} {\bibfield
   {journal} {\bibinfo  {journal} {Phys. Rev. D}\ }\textbf {\bibinfo {volume}
  {82}},\ \bibinfo {pages} {044051} (\bibinfo {year} {2010})},\ \Eprint
  {https://arxiv.org/abs/1005.2266} {arXiv:1005.2266 [gr-qc]} \BibitemShut
  {NoStop}%
\bibitem [{\citenamefont {Pound}\ \emph {et~al.}(2020)\citenamefont {Pound},
  \citenamefont {Wardell}, \citenamefont {Warburton},\ and\ \citenamefont
  {Miller}}]{Pound:2019lzj}%
  \BibitemOpen
  \bibfield  {author} {\bibinfo {author} {\bibfnamefont {A.}~\bibnamefont
  {Pound}}, \bibinfo {author} {\bibfnamefont {B.}~\bibnamefont {Wardell}},
  \bibinfo {author} {\bibfnamefont {N.}~\bibnamefont {Warburton}},\ and\
  \bibinfo {author} {\bibfnamefont {J.}~\bibnamefont {Miller}},\ }\bibfield
  {title} {\bibinfo {title} {{Second-Order Self-Force Calculation of
  Gravitational Binding Energy in Compact Binaries}},\ }\href
  {https://doi.org/10.1103/PhysRevLett.124.021101} {\bibfield  {journal}
  {\bibinfo  {journal} {Phys. Rev. Lett.}\ }\textbf {\bibinfo {volume} {124}},\
  \bibinfo {pages} {021101} (\bibinfo {year} {2020})},\ \Eprint
  {https://arxiv.org/abs/1908.07419} {arXiv:1908.07419 [gr-qc]} \BibitemShut
  {NoStop}%
\bibitem [{\citenamefont {Lewis}\ \emph {et~al.}(2025)\citenamefont {Lewis},
  \citenamefont {Kakehi}, \citenamefont {Pound},\ and\ \citenamefont
  {Tanaka}}]{LKPT25}%
  \BibitemOpen
  \bibfield  {author} {\bibinfo {author} {\bibfnamefont {J.}~\bibnamefont
  {Lewis}}, \bibinfo {author} {\bibfnamefont {T.}~\bibnamefont {Kakehi}},
  \bibinfo {author} {\bibfnamefont {A.}~\bibnamefont {Pound}},\ and\ \bibinfo
  {author} {\bibfnamefont {T.}~\bibnamefont {Tanaka}},\ }\bibfield  {title}
  {\bibinfo {title} {{Post-adiabatic dynamics and waveform generation in
  self-force theory: an invariant pseudo-Hamiltonian framework}},\ }\href@noop
  {} {\  (\bibinfo {year} {2025})},\ \Eprint {https://arxiv.org/abs/2507.08081}
  {arXiv:2507.08081 [gr-qc]} \BibitemShut {NoStop}%
\end{thebibliography}%
\end{document}